\documentclass[12pt]{article}
\usepackage{epsfig}
\usepackage{amsmath}
\usepackage{hhline}
\usepackage{amssymb}
\usepackage{times}
\usepackage{cite}
\usepackage{multirow}
\usepackage{multicol}
\usepackage{array}
\usepackage{longtable}

\newlength{\dinwidth}
\newlength{\dinmargin}
\begin{document}  
\newcommand{\pom}{{I\!\!P}}
\newcommand{\reg}{{I\!\!R}}
\newcommand{\slowpi}{\pi_{\mathit{slow}}}
\newcommand{\fiidiii}{F_2^{D(3)}}
\newcommand{\fiidiiiarg}{\fiidiii\,(\beta,\,Q^2,\,x)}
\newcommand{\n}{1.19\pm 0.06 (stat.) \pm0.07 (syst.)}
\newcommand{\nz}{1.30\pm 0.08 (stat.)^{+0.08}_{-0.14} (syst.)}
\newcommand{\fiidiiiful}{F_2^{D(4)}\,(\beta,\,Q^2,\,x,\,t)}
\newcommand{\fiipom}{\tilde F_2^D}
\newcommand{\ALPHA}{1.10\pm0.03 (stat.) \pm0.04 (syst.)}
\newcommand{\ALPHAZ}{1.15\pm0.04 (stat.)^{+0.04}_{-0.07} (syst.)}
\newcommand{\fiipomarg}{\fiipom\,(\beta,\,Q^2)}
\newcommand{\pomflux}{f_{\pom / p}}
\newcommand{\nxpom}{1.19\pm 0.06 (stat.) \pm0.07 (syst.)}
\newcommand {\gapprox}
   {\raisebox{-0.7ex}{$\stackrel {\textstyle>}{\sim}$}}
\newcommand {\lapprox}
   {\raisebox{-0.7ex}{$\stackrel {\textstyle<}{\sim}$}}
\def\gsim{\,\lower.25ex\hbox{$\scriptstyle\sim$}\kern-1.30ex%
\raise 0.55ex\hbox{$\scriptstyle >$}\,}
\def\lsim{\,\lower.25ex\hbox{$\scriptstyle\sim$}\kern-1.30ex%
\raise 0.55ex\hbox{$\scriptstyle <$}\,}
\newcommand{\pomfluxarg}{f_{\pom / p}\,(x_\pom)}
\newcommand{\dsf}{\mbox{$F_2^{D(3)}$}}
\newcommand{\dsfva}{\mbox{$F_2^{D(3)}(\beta,Q^2,x_{I\!\!P})$}}
\newcommand{\dsfvb}{\mbox{$F_2^{D(3)}(\beta,Q^2,x)$}}
\newcommand{\dsfpom}{$F_2^{I\!\!P}$}
\newcommand{\gap}{\stackrel{>}{\sim}}
\newcommand{\lap}{\stackrel{<}{\sim}}
\newcommand{\fem}{$F_2^{em}$}
\newcommand{\tsnmp}{$\tilde{\sigma}_{NC}(e^{\mp})$}
\newcommand{\tsnm}{$\tilde{\sigma}_{NC}(e^-)$}
\newcommand{\tsnp}{$\tilde{\sigma}_{NC}(e^+)$}
\newcommand{\st}{$\star$}
\newcommand{\sst}{$\star \star$}
\newcommand{\ssst}{$\star \star \star$}
\newcommand{\sssst}{$\star \star \star \star$}
\newcommand{\tw}{\theta_W}
\newcommand{\sw}{\sin{\theta_W}}
\newcommand{\cw}{\cos{\theta_W}}
\newcommand{\sww}{\sin^2{\theta_W}}
\newcommand{\cww}{\cos^2{\theta_W}}
\newcommand{\trm}{m_{\perp}}
\newcommand{\trp}{p_{\perp}}
\newcommand{\trmm}{m_{\perp}^2}
\newcommand{\trpp}{p_{\perp}^2}
\newcommand{\alp}{\alpha_s}

\newcommand{\alps}{\alpha_s}
\newcommand{\sqrts}{$\sqrt{s}$}
\newcommand{\LO}{$O(\alpha_s^0)$}
\newcommand{\Oa}{$O(\alpha_s)$}
\newcommand{\Oaa}{$O(\alpha_s^2)$}
\newcommand{\PT}{p_{\perp}}
\newcommand{\JPSI}{J/\psi}
\newcommand{\sh}{\hat{s}}
\newcommand{\uh}{\hat{u}}
\newcommand{\MP}{m_{J/\psi}}
\newcommand{\PO}{I\!\!P}
\newcommand{\xbj}{x}
\newcommand{\xpom}{x_{\PO}}
\newcommand{\ttbs}{\char'134}
\newcommand{\xpomlo}{3\times10^{-4}}  
\newcommand{\xpomup}{0.05}  
\newcommand{\dgr}{^\circ}
\newcommand{\pbarnt}{\,\mbox{{\rm pb$^{-1}$}}}
\newcommand{\gev}{\,\mbox{GeV}}
\newcommand{\WBoson}{\mbox{$W$}}
\newcommand{\fbarn}{\,\mbox{{\rm fb}}}
\newcommand{\fbarnt}{\,\mbox{{\rm fb$^{-1}$}}}
\newcommand{\dsdx}[1]{$d\sigma\!/\!d #1\,$}
\newcommand{\eV}{\mbox{e\hspace{-0.08em}V}}
%
%
\newcommand{\qsq}{\ensuremath{Q^2} }
\newcommand{\gevsq}{\ensuremath{\mathrm{GeV}^2} }
\newcommand{\et}{\ensuremath{E_t^*} }
\newcommand{\rap}{\ensuremath{\eta^*} }
\newcommand{\gp}{\ensuremath{\gamma^*}p }
\newcommand{\dsiget}{\ensuremath{{\rm d}\sigma_{ep}/{\rm d}E_t^*} }
\newcommand{\dsigrap}{\ensuremath{{\rm d}\sigma_{ep}/{\rm d}\eta^*} }

\newcommand{\dstar}{\ensuremath{D^*}}
\newcommand{\dstarp}{\ensuremath{D^{*+}}}
\newcommand{\dstarm}{\ensuremath{D^{*-}}}
\newcommand{\dstarpm}{\ensuremath{D^{*\pm}}}
\newcommand{\zDs}{\ensuremath{z(\dstar )}}
\newcommand{\Wgp}{\ensuremath{W_{\gamma p}}}
\newcommand{\ptds}{\ensuremath{p_t(\dstar )}}
\newcommand{\etads}{\ensuremath{\eta(\dstar )}}
\newcommand{\ptj}{\ensuremath{p_t(\mbox{jet})}}
\newcommand{\ptjn}[1]{\ensuremath{p_t(\mbox{jet$_{#1}$})}}
\newcommand{\etaj}{\ensuremath{\eta(\mbox{jet})}}
\newcommand{\detadsj}{\ensuremath{\eta(\dstar )\, \mbox{-}\, \etaj}}

\def\Journal#1#2#3#4{{#1} {\bf #2} (#3) #4}
\def\NCA{\em Nuovo Cimento}
\def\NIM{\em Nucl. Instrum. Methods}
\def\NIMA{{\em Nucl. Instrum. Methods} {\bf A}}
\def\NPB{{\em Nucl. Phys.}   {\bf B}}
\def\PLB{{\em Phys. Lett.}   {\bf B}}
\def\PRL{\em Phys. Rev. Lett.}
\def\PRD{{\em Phys. Rev.}    {\bf D}}
\def\ZPC{{\em Z. Phys.}      {\bf C}}
\def\EJC{{\em Eur. Phys. J.} {\bf C}}
\def\CPC{\em Comp. Phys. Commun.}

\begin{titlepage}

\noindent
\begin{flushleft}
DESY 10-043\hfill ISSN 0418-9833\\
March 2010
\end{flushleft}

\vspace*{2cm}

\begin{center}
\begin{Large}

{\bf Diffractive Dijet Photoproduction \\ 
in {\boldmath $ep$} Collisions at HERA}

\vspace{2cm}

H1 Collaboration

\end{Large}
\end{center}

\vspace{2cm}

\begin{abstract}
Measurements are presented of single and double-differential dijet cross 
sections in diffractive photoproduction based on a data sample
with an integrated luminosity of 
\linebreak
$47 \ {\rm pb^{-1}}$. The events
are of the type $ep \rightarrow eXY$, where the 
hadronic system $X$ contains at 
least two jets and is separated by a large rapidity gap from the
system $Y$, which consists of a leading proton or 
low-mass proton excitation. The dijet cross sections are compared 
with QCD calculations at next-to-leading order and with a Monte Carlo 
model based on leading order matrix elements with parton showers. The 
measured cross sections are smaller than those obtained from the
next-to-leading order calculations by a factor of about $0.6$. This 
suppression factor has no significant dependence on the fraction
$x_{\gamma}$ of the photon four-momentum entering the hard subprocess. 
Ratios of the diffractive to 
the inclusive dijet cross sections are measured
for the first time and are compared with Monte Carlo 
models.
\end{abstract}

\vspace{2cm}

\begin{center}
Submitted to \EJC
\end{center}

\end{titlepage}


\noindent
F.D.~Aaron$^{5,49}$,           
C.~Alexa$^{5}$,                
V.~Andreev$^{25}$,             
S.~Backovic$^{30}$,            
A.~Baghdasaryan$^{38}$,        
E.~Barrelet$^{29}$,            
W.~Bartel$^{11}$,              
K.~Begzsuren$^{35}$,           
A.~Belousov$^{25}$,            
J.C.~Bizot$^{27}$
% ...

\bigskip{\it
\noindent
% ...
$ ^{5}$ National Institute for Physics and Nuclear Engineering (NIPNE) ,
          Bucharest, Romania$^{ m}$ \\
% ...
$ ^{11}$ DESY, Hamburg, Germany \\
% ...
$ ^{25}$ Lebedev Physical Institute, Moscow, Russia$^{ e}$ \\
% ...
$ ^{27}$ LAL, Universit\'e Paris-Sud, CNRS/IN2P3, Orsay, France \\
% ...
$ ^{29}$ LPNHE, Universit\'e Pierre et Marie Curie Paris 6,
          Universit\'e Denis Diderot Paris 7, CNRS/IN2P3, Paris, France \\
 $ ^{30}$ Faculty of Science, University of Montenegro,
          Podgorica, Montenegro$^{ e}$ \\
% ...
$ ^{35}$ Institute of Physics and Technology of the Mongolian
          Academy of Sciences, Ulaanbaatar, Mongolia \\
% ...
$ ^{38}$ Yerevan Physics Institute, Yerevan, Armenia \\
% ...
$ ^{49}$ Also at Faculty of Physics, University of Bucharest,
          Bucharest, Romania \\
% ...
$ ^e$ Supported by the Deutsche Forschungsgemeinschaft \\
% ...
$ ^m$ Supported by the Romanian National Authority for Scientific Research
      under the contract PN 09370101 \\
}

\newpage

\section{Introduction}
Quantum Chromodynamics (QCD), as the gauge field theory of the strong 
interaction, reliably predicts scattering 
cross sections 
involving short distance partonic interactions. However, the 
vast majority of
hadron-hadron scatterings take place through long-distance strong interactions,
where no hard scales are present and perturbative QCD calculations are not
possible. Prominent among these soft interactions are 
diffractive processes, in which the interacting hadrons remain intact or 
dissociate into low mass hadronic systems 
via an exchange which has vacuum quantum
numbers, often referred to as a `pomeron' \cite{Diff:review}. 
Following the observation of diffractive $p\bar{p}$ collisions 
in which a hard scale is provided by high transverse momentum 
jets \cite{UA8}, it has become possible to describe some classes 
of diffractive processes in terms of partonic 
interactions \cite{IS}.
More recently, diffractive deep-inelastic scattering (DDIS)
processes at HERA \cite{HERA:Diff}, of the type $ep \rightarrow eXp$, have 
been studied in detail and have led to a new level of understanding of
the properties and structure of the diffractive exchange. Developing this
microscopic description of diffraction 
in terms of QCD and parton dynamics is a step
towards a more complete understanding of the strong interaction.

In the framework of a collinear factorisation theorem \cite{Collins} 
for hard scattering in semi-inclusive processes such as
DDIS, diffractive parton distribution functions (DPDFs) 
may be defined. The DPDFs have similar
properties to the standard parton distribution functions (PDFs) of the 
proton, but with the constraint that there be a leading proton present 
in the final state. 
This condition may be satisfied equivalently by the experimental signatures
of either a leading proton \cite{H1:FPS,zeus:f2d} or the presence of a large 
gap in the rapidity distribution of final state hadrons, separating an
unobserved outgoing proton from the remainder of the hadronic final 
state \cite{H1:LRG,zeus:f2d}. In various extractions using recent
DDIS data \cite{H1:LRG,Matthias,ZEUS:DPDF,DPDFs}, 
the DPDFs have been found to 
be dominated by 
gluons. To good approximation they
exhibit a `proton vertex factorisation' property, 
whereby they vary only in normalisation
with the four-momentum of the final state proton, the normalisation 
being well modelled using Regge phenomenology \cite{Regge}. 

Given a knowledge of the DPDFs, perturbative QCD calculations are applicable
to other DDIS observables. Such 
calculations have been successful in the prediction of
jet \cite{HERA:jet,DIS:jet,Matthias} and 
heavy quark \cite{DIS:hf} production in DDIS at HERA. In both cases,
next-to-leading order (NLO) QCD predictions using the DPDFs
from \cite{H1:LRG} describe the measured cross sections well. 
However, as has long been anticipated \cite{fac:break,Collins}, 
DPDF-based predictions for hard
diffractive processes such as dijet production
in $p\bar{p}$ scattering fail by around an order of magnitude to 
describe the data~\cite{TevatronHERA,KKTevatron}. 
This factorisation breaking is 
generally attributed to absorptive corrections, corresponding to
the destruction of the rapidity gap due to
multiple interactions within a single event. 
Such effects are possible in 
$p\bar{p}$ scattering, where a beam remnant is present, in contrast to the
electron scattering case in DDIS at HERA. 
A diversity of models of the absorptive corrections has 
developed \cite{Gap:Survival,Kaidalov}, several of which
reproduce the approximate  
$ 10\%$ `rapidity gap survival probability' observed in 
single diffractive $p\bar{p}$ scattering.

The issues of DPDF applicability and rapidity gap survival can be studied
in $ep$ collisions at HERA in hard diffractive 
`photoproduction', where the virtuality 
$Q^2$ of the 
exchange photon is close to zero. 
Under these circumstances, the photon can develop 
an effective partonic structure via $\gamma \rightarrow q\bar{q}$ fluctuations
and further subsequent splittings \cite{hard:gammap}. 
In a leading order 
picture, there 
are thus two classes of hard photoproduction: 
`direct' interactions,
where the photon enters the hard scatter as a structureless object and
`resolved' interactions, where 
the photon interacts via its partonic structure and 
only a fraction $x_\gamma$ 
of its four-momentum participates in the hard subprocess.
Resolved photoproduction interactions can be further divided into
a `hadron-like' contribution and an `anomalous' or `point-like' 
contribution, the latter arising from the inhomogeneous term in the DGLAP
equation for the photon \cite{witten}.
Interactions involving the hadron-like component resemble 
hadron-hadron scattering to
a large extent and 
are therefore widely expected to exhibit gap destruction effects.
The rapidity gap survival
probability for these hadron-like processes has
been estimated in a phenomenological 
model to be $0.34$ \cite{kkmr,kkmr:new}.
The point-like contribution to photon structure is expected to
be subject to smaller absorptive corrections than the hadron-like
part \cite{klasen}. In a recent model \cite{kkmr:new}
a survival probability of around $0.7 - 0.8$ was obtained
for diffractive dijet photoproduction,
depending slightly on the jet transverse energies ($E_T^{\rm jet}$).

Previous H1 measurements of diffractive dijet 
photoproduction \cite{HERA:jet,gp:jet} have found 
cross sections to be smaller than NLO theoretical predictions, 
suggesting
rapidity gap survival probabilities of around $0.5$ with
little dependence on $x_\gamma$. A recent ZEUS 
measurement at somewhat larger $E_T^{\rm jet}$ \cite{ZEUSphp} 
yielded a larger survival probability, compatible with
unity. It has been proposed that this apparent 
discrepancy may be resolved if the rapidity gap survival probability
depends on the scale of the hard interaction, an idea which is 
supported to some extent by data \cite{gp:jet,ZEUSphp,karel:DIS08}. 
Neither H1 nor ZEUS data provide any evidence for the expected 
$x_\gamma$ dependence of the rapidity gap survival probability. 

A measurement of the ratio of the diffractive to the inclusive 
dijet photoproduction cross sections 
was proposed in~\cite{kkmr,klasen,Klasen0.34} as a means of evaluating 
the gap survival probability.
This ratio is expected to be relatively insensitive to the 
model of the photon parton densities and also
offers cancellations of experimental systematics and higher order QCD 
corrections.
A similar ratio was measured by the CDF
collaboration \cite{TevatronHERA}
as a means of extracting effective $p \bar{p}$
DPDFs for comparison with HERA predictions and assessment of gap survival
probabilities.

This paper reports diffractive dijet photoproduction 
cross section measurements based on
a positron-proton scattering
data sample with luminosity about a factor three larger 
than that previously published by H1 \cite{gp:jet}. The larger sample 
makes double-differential measurements possible, giving greater detail
on the dynamics of gap survival and allowing studies of the 
correlations between the 
kinematic variables. The hypothesis of an $E_T^{\rm jet}$ dependent 
rapidity gap survival probability is tested.
The ratio of the diffractive to the 
inclusive dijet
photoproduction cross sections is also extracted for the first time.

\section{Kinematic Variables}
\label{sec:kin}

\begin{figure}[hhh]
\begin{center}
\epsfig{file=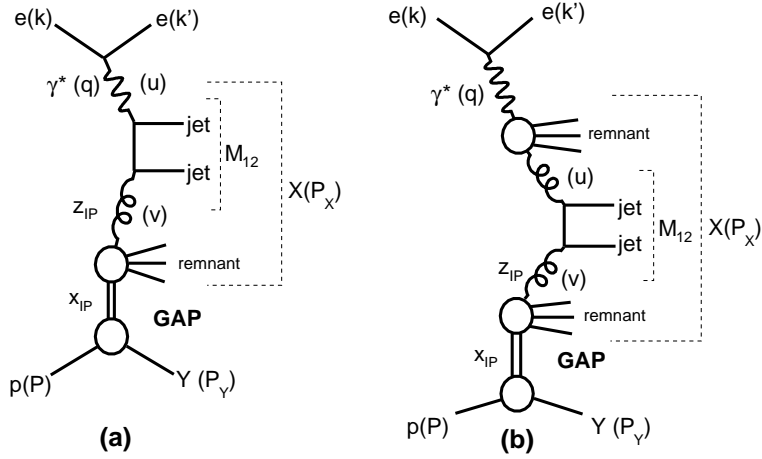 ,width=10cm,bbllx=0pt,bblly=5pt,bburx=555pt,bbury=338pt,angle=0,clip=}
\caption{Leading order diagrams for diffractive dijet photoproduction at 
HERA. Diagrams (a) and (b) are examples of direct and resolved photon 
interactions, respectively.}
\label{fig:dirresfeyn}
\end{center}
\end{figure}

Figures~\ref{fig:dirresfeyn} (a) and (b) show leading order 
examples of direct and resolved 
diffractive dijet production.
Denoting the four-vectors of the incoming positron, the incoming 
proton and the exchanged photon as $k$, $P$ and $q$, respectively,
the standard DIS kinematics can be described in terms of the invariants
\begin{eqnarray}
s \equiv (k+P)^{2} \hspace*{1.3cm} 
Q^2 \equiv -q^{2} \hspace*{1.3cm} 
y \equiv {{q\cdot P}\over{k\cdot P}} \hspace*{1.3cm} 
x \equiv {{Q^{2}}\over{2\,q\cdot P}} \ .
\label{sqyx}
\end{eqnarray}
Here,
$s$ is the square of the total centre of mass energy of the collision, 
$Q^{2}$ is the photon virtuality, $y$ 
is the scattered positron inelasticity and $x$ is the 
fraction of the proton four-momentum carried by the quark coupling 
to the photon. 
These variables are related through $Q^{2} = s\,x\,y$ and to the 
invariant mass $W$ of the photon-proton system by
\begin{eqnarray}
W \equiv \sqrt{(q+P)^{2}} \approx \sqrt{ys - Q^2} \ .
\label{w}
\end{eqnarray}

Defining $P_{X}$ and $P_{Y}$ to be the four-vectors of the two 
distinct final state
systems, where $Y$ may be either a proton or a low mass proton excitation, 
the diffractive
kinematics are described by the variables
\begin{eqnarray}
M_{X}^2 \equiv P_{X}^{2} \hspace*{1.3cm} 
M_{Y}^2 \equiv P_{Y}^{2} \hspace*{1.3cm} 
t\equiv (P-P_{Y})^{2} \hspace*{1.3cm} 
x_{\pom} \equiv {{q\cdot (P-P_{Y})}\over{q\cdot P}} \ .
\label{mxmytxpom}
\end{eqnarray}
Here, $M_{X}$ and $M_{Y}$ are the invariant masses of the 
systems $X$ and $Y$, $t$ is the 
squared four-momentum 
transfer at the proton vertex and 
$x_{\pom}$ is the fraction of the
longitudinal momentum of the proton transferred to 
the system $X$. 

With $u$ and $v$ being the four-momenta of the particles entering the hard
scatter from the photon and proton, respectively ($u = q$ in the
direct photon case), the invariant mass $M_{12}$ of the dijet system and the 
fractional photon ($x_{\gamma}$) and pomeron ($z_{\pom}$) 
longitudinal momenta entering the hard subprocess can be expressed as
\begin{eqnarray}
M_{12} \equiv \sqrt{(u+v)^{2}} \hspace*{1.3cm} 
x_{\gamma} \equiv {{P\cdot u}\over{P\cdot q}} \hspace*{1.3cm} 
z_{\pom} \equiv {{q\cdot v}\over{q\cdot (P-P_{Y})}} \ .
\label{m12,zg,zpom}
\end{eqnarray}

\section{Theory and Models}
\label{sec:theory}
\subsection{Diffractive Dijet 
Photoproduction in the Factorisation Approach} 
Dijet electroproduction cross sections for $Q^2 \rightarrow 0$
can be calculated in a fixed order 
QCD approach, assuming QCD collinear factorisation and neglecting any 
rapidity gap destruction effects, as a convolution of partonic 
cross sections, photon PDFs and DPDFs, according to 
\begin{eqnarray}
\label{eq:factorised cross section}
\nonumber
d\sigma(ep\rightarrow e + 2 \, {\rm jets} + X^{\prime} + Y) = 
\sum_{i,j} \int dy \,\, f_{\gamma/e} (y) \int dx_{\gamma} 
f_{j/\gamma}(x_{\gamma}, \mu_{F}^{2}) \ \otimes \\  
\int dt \int dx_{\pom} \int dz_{\pom} \ d\hat{\sigma}(ij \rightarrow 2 \,
{\rm jets}) \
f_{i/p}^{D}(z_{\pom}, \mu_{F}^{2}, x_{\pom}, t)\ .
\end{eqnarray} 
Here, the hadronic system $X^{\prime}$ corresponds to the remainder of
the system $X$ after removing the two jets. The sum runs over all partons 
$i$ and $j$ that contribute, $f_{\gamma/e}$ is the 
equivalent flux of photons 
that emerge from the incoming lepton \cite{wwa}
and $f_{j/\gamma}$ are the photon 
PDFs ($f_{j/\gamma} = \delta(1-x_{\gamma})$ in the direct
photon case). The hard partonic cross sections 
are denoted $\hat{\sigma}$, $f_{i/p}^{D}$ are 
the DPDFs of the proton and $\mu_{F}$ is the factorisation scale. 

In this analysis, 
the GRV HO~\cite{GRV} parton densities are used to describe the 
structure of the resolved photon. 
The H1 2006 Fit B set is used for the DPDFs,
obtained by the H1 collaboration 
in fits to inclusive DDIS data \cite{H1:LRG}.
In the poorly constrained large $z_{\pom}$ region,
previous DDIS final state data \cite{Matthias,gp:jet,DIS:hf} 
have shown a clear 
preference for these DPDFs over the Fit A set from \cite{H1:LRG}.
Further DPDF sets from H1 (H1 2007 Fit Jets) \cite{Matthias}
and ZEUS (ZEUS DPDF SJ) \cite{zeus:f2d,ZEUS:DPDF} in which
dijet data from DDIS are used to improve the sensitivity 
to the large $z_{\pom}$ region, are also considered. 
All of these DPDF sets assume proton vertex factorisation,
such that
\begin{eqnarray}
\label{eq:regge factorization}
f_{i/p}^{D}(z_{\pom}, \mu_{F}^{2}, x_{\pom}, t) = f_{\pom/p}(x_{\pom},\,t) f_{i/\pom}(z_{\pom}, \,\mu_{F}^{2}) \ .
\end{eqnarray} 
In the interpretation illustrated in figure~\ref{fig:dirresfeyn}, 
$f_{\pom/p}(x_{\pom},t)$ may be considered as a pomeron 
flux, parameterised using Regge phenomenology in the DPDF sets
used here. The
 $f_{i/\pom}$ factor then represents the parton densities of the pomeron. 
At relatively large values of $x_{\pom} \, \gapprox \, 0.01$ a small 
contribution from a sub-leading meson exchange 
is required in the DPDF fits. This is
taken into account by adding a second term of the same form as
equation~\ref{eq:regge factorization}, but with different 
parton densities and a flux factor which is suppressed as 
$x_{\pom} \rightarrow 0$.

\subsection{Next-to-leading Order Parton Level QCD Calculations}
\label{nlo}

The dijet electroproduction 
cross sections (equation~\ref{eq:factorised cross section})
are calculated at NLO of QCD using the program of 
Frixione {\it et al}. (FR) \cite{FR} adapted for diffractive applications as 
described in \cite{seb:thesis,gp:jet}. 
In the FR program, the renormalisation and 
factorisation scales are set to be equal and both are taken here from 
the leading jet transverse energy, 
i.e $\mu_{R} = \mu_{F} = E_{T}^{\rm jet1}$. The NLO calculations are 
performed with the number of
flavours fixed to 5 and the 
QCD scale parameter set to $\Lambda_{5} = 0.228 \ {\rm GeV}$,
corresponding to a 2-loop $\alpha_{s}(M_{Z})=0.118$. 
The sensitivity of the calculated cross sections to the chosen
$\mu_{R}$ and $\mu_{F}$ values is studied by varying both scales 
simultaneously by factors of $0.5$ and $2$. 
The NLO calculations were cross-checked with the program written by
Klasen and Kramer \cite{klasen}, which yields consistent 
results \cite{karel:DIS08,karel:thesis}. 


\subsection{Monte Carlo Simulations}

\subsubsection{Corrections to the Data}
\label{sec:MC}

A Monte Carlo (MC) simulation is
used to correct the data for detector effects in obtaining 
cross sections at the level of stable hadrons. 
All MC samples are passed through a detailed simulation of
the H1 detector response based on the GEANT program \cite{geant}
and are subjected to the same reconstruction and analysis 
algorithms as are used for the data.

Diffractive dijet photoproduction events are generated 
using the RAPGAP MC generator~\cite{Rapgap}
in the 
range $Q^{2} < 0.01$ GeV$^{2}$ with the minimum 
transverse momentum of the partons 
entering the hard subprocess
set to
$\hat{p}_{T}^{min} = 2~\mbox{GeV}$. RAPGAP
is based on leading order QCD
matrix elements with DGLAP parton showers.
The H1 2006 fit B 
DPDFs and the 
GRV-G LO photon parton densities \cite{GRV} are used at a 
factorisation scale given by $E_{T}^{\rm jet1}$. 
A sub-leading meson exchange is included in the DPDF simulation,
though its contribution is smaller than $5\%$ in the kinematic
range covered here.

The selection 
of diffractive events using the rapidity gap method
(section~\ref{sec:selection}) yields a sample
which is dominated by elastically scattered protons, but which also 
contains an admixture of events in which the proton dissociates to
low $M_Y$ states. The measurement is corrected to the 
region $M_Y < 1.6 \ {\rm GeV}$ and $|t| < 1 \ {\rm GeV^2}$
(section~\ref{sec:xsec}), 
using MC samples generated using the 
DIFFVM \cite{DIFFVM} program, 
with and without proton dissociation, following the
method described in \cite{H1:LRG}. 

In order to estimate the small ($\sim 2 \%$)
contribution to the data sample from 
non-diffractive events which pass the diffractive selection,
the PYTHIA MC generator~\cite{Pythia} is used in 
photoproduction mode ($Q^{2} < 0.01~\mbox{GeV}^{2}$).
The PYTHIA model is also used to correct 
the inclusive photoproduction
dijet measurement
for detector effects
when extracting
the ratio of diffractive to inclusive cross sections. 
To estimate the model uncertainties on the corrections to the
inclusive dijet cross section,
a further MC
sample is obtained using the HERWIG generator \cite{Herwig}. 
Multiple interaction models are included in both PYTHIA and HERWIG
as described in section~\ref{sec:MChad}. 
With these settings, PYTHIA provides a good description of 
the shapes of the uncorrected 
inclusive dijet distributions,
with a normalisation slightly larger than that of the data. 
The HERWIG MC underestimates the normalisation of 
the cross section by about a factor of two, but gives an acceptable
description of the shapes of the measured distributions. 

\subsubsection{Corrections to Theoretical Models}
\label{sec:MChad}

For comparison with the diffractive measurements, it is necessary to 
convert the calculated NLO parton-level cross
sections to the level of stable hadrons by evaluating effects due to
initial and final state parton
showering, fragmentation, hadronisation and the influence of beam remnants.
The RAPGAP MC is used to compute the required 
`hadronisation correction' factors to the  
diffractive dijet calculations. These factors are defined
for each measured data point by
\begin{eqnarray}
\label{hadrcorr}
1 + \delta_{hadr.} = 
{{\sigma_{\rm dijet}^{\rm hadron}}\over{\sigma_{\rm dijet}^{\rm parton}}} \ .
\end{eqnarray}
They reduce the predicted cross sections
by typically $15\%$ and are given for each data point in tables~\ref{table1D} 
and~\ref{table2D}.
The shape of the $x_{\gamma}$ distribution is
most strongly affected by the hadronisation 
corrections,
the main effect being the migration of some
direct photon interactions, for which
the cross section is large, to lower $x_{\gamma}$ values,
substantially increasing the prediction in the interval
$0.6 < x_{\gamma} < 0.8$.

Resolved photon interactions in
inclusive dijet production \cite{MI} are poorly described 
by NLO calculations unless a model of 
multiple interactions (MI) within a single event
is included in addition to the low transverse momentum 
beam remnant, QCD radiation 
and hadronisation contributions which are
simulated in standard MC models.
The PYTHIA MC generator is used 
to investigate the influence of MIs 
on inclusive dijet cross sections
and hence on the diffractive-to-inclusive ratios.
Several tunes of the
PYTHIA model for multiple hard parton-parton
scatterings are available.
In the version used here (tune A in \cite{Pythia}), 
both the proton and the resolved photon have  
double-Gaussian matter 
distributions. The minimum transverse momentum 
down to which secondary partonic scatterings are calculated
depends on the impact parameter of the collision and is governed by a
regularisation scale $p_{\perp0} = 1.2 \ {\rm GeV}$ \cite{Pythia}.

Multiple 
hard partonic interactions
are not simulated in the HERWIG MC. Instead, an empirically motivated 
soft multiple interaction model \cite{Herwig} is used.
The probability of such activity in a given event is set to $0.25$.

\section{Experimental Procedure}
\label{sec:procedure}

\subsection{The H1 Detector}
\label{sec:H1 detector}

A detailed description of the H1 detector can be found 
elsewhere \cite{H1detector}. 
Here, a brief account is given of the detector components 
most relevant to the present analysis.
The H1 coordinate system
is defined such that the origin is at the nominal $ep$ interaction point and
the polar angle
$\theta = 0^\circ$ and the positive $z$ axis correspond to the 
direction of the outgoing proton beam.
The region $\theta < 90^\circ$, which has
positive pseudorapidity $\eta = - \ln \tan
\theta /2$, is referred to as the `forward' hemisphere.

The {\it ep} interaction point in H1 is surrounded by a central tracking 
region, which includes silicon strip detectors as well as two 
large concentric drift chambers. These chambers cover a pseudorapidity region 
of $-1.5 < \eta < 1.5$ and have a transverse momentum resolution of 
$\sigma(P_{T})/P_{T} = 0.006P_{T}/{\rm GeV} \oplus 0.02$. They also  
provide triggering information. The central tracking detectors are surrounded 
by a finely segmented liquid argon (LAr) sampling calorimeter covering 
$-1.5 < \eta < 3.4$. Its resolution is
$\sigma/E = 0.11/\sqrt{E/{\rm GeV}} \oplus 0.01$
for electrons and photons and
$\sigma/E = 0.50/\sqrt{E/{\rm GeV}} \oplus 0.02$ 
for hadrons, as measured in test beams \cite{testbeam}. 
The central tracker and LAr calorimeter are placed inside 
a large superconducting solenoid, which produces a uniform magnetic 
field of $1.16$ T. The backward region $-4 < \eta < -1.4$ is covered by 
a lead-scintillating fibre calorimeter (SpaCal) with electromagnetic and
hadronic sections. 

Information from the central tracker and the LAr and SpaCal calorimeters 
is combined using an energy flow algorithm to obtain 
the hadronic final state (HFS) \cite{HFS:alg}.
The hadronic energy scale 
is known to $3\%$ 
for this analysis \cite{karel:thesis}.

Photoproduction events are selected by tagging positrons scattered through
very small angles, corresponding to quasi-real photon emission,  
using a crystal \v{C}erenkov calorimeter at $z=-33~\mbox{m}$
(electron tagger).
The luminosity is measured via the Bethe-Heitler bremsstrahlung 
process $ep \rightarrow ep\gamma$, the final state photon being 
detected in another crystal calorimeter at $z=-103~\mbox{m}$.

A set of drift chambers around $z = 6.5~\mbox{m}$
comprise the forward muon detector (FMD). The proton remnant tagger (PRT)
is a set of scintillators surrounding the beam pipe at 
$z=26~\mbox{m}$. These detectors, used together
with the most forward part of the LAr, are efficient in the 
identification of very forward energy flow and are used to select events
with large rapidity gaps near to the outgoing proton 
direction.

\subsection{Event Selection, Kinematic and Jet Reconstruction}
\label{sec:selection}

The analysis is based on a sample of integrated luminosity 
$47 \ {\rm pb^{-1}}$,
collected by H1 in 1999 and 2000 with proton and 
positron beam energies of $920 \ {\rm GeV}$ and $27.5 \ {\rm GeV}$, 
respectively. The events
are triggered on the basis of a scattered positron signal in the electron
tagger and at least 
three high transverse momentum tracks in the drift chambers
of the central tracker. 

The event inelasticity $y$ 
and hence the invariant mass $W$ of the photon-proton
system are reconstructed using the scattered positron energy
$E_{e'}$ measured in the electron tagger according to
\begin{eqnarray}
y = 1 - {{E_{e'}}\over{E_{e}^{beam}}} \hspace*{1.3cm}
W = \sqrt{s \, y} \ ,
\end{eqnarray}
where $E_{e}^{beam}$ is the positron beam energy.
The geometric acceptance of the electron tagger limits the measurement to 
$Q^{2} < 0.01~\mbox{GeV}^2$ and intermediate values of $y$. 

The reconstructed hadronic final state objects (section~\ref{sec:H1 detector})
are subjected to the $k_{T}$ longitudinally invariant
jet algorithm \cite{Catani}, applied 
in the laboratory frame with parameters $R = 1$ and 
$E_{T}^{\rm jet,min} = 2.5\,\,\mbox{GeV}$. 
To facilitate comparisons with the NLO calculations, 
different cuts are placed on the transverse energies $E_{T}^{\rm jet1}$ and
$E_{T}^{\rm jet2}$ of the leading and next-to-leading jets, 
respectively. 
As well as these variables, the jet properties are studied in terms
of the variables 
\begin{eqnarray}
{\left| \Delta \eta^{\rm jets} \right|} =  
{\left| \eta^{\rm jet1} - \eta^{\rm jet2} \right|} \hspace*{1.3cm}
\left< \eta^{\rm jets} \right> = 
{{1}\over{2}} \left( \eta^{\rm jet1} + \eta^{\rm jet2} \right) \ ,
\end{eqnarray}
obtained from the laboratory frame pseudorapidities of the jet axes.
With $J^{(1)}$ and $J^{(2)}$ denoting the four-momenta of the two jets,
hadron level estimators of the dijet invariant mass 
and of $x_\gamma$ are obtained from 
\begin{eqnarray}
M_{12} & = & \sqrt{2 \, J^{(1)} \cdot J^{(2)}} \hspace*{1.3cm} 
x_{\gamma}^{\rm jets} = {{\Sigma_{jets} (E _{i} - P_{z,i})} 
\over {\Sigma_{HFS} (E_{i} - P_{z,i})}} \ ,
\end{eqnarray}
where the sums labelled `{\it HFS}' and `{\it jets}' 
run over all hadronic final state 
objects and those included in the jets, respectively.

The diffractive event selection is based on the presence of a large 
forward rapidity gap. The pseudorapidity of 
the most forward cluster in the 
LAr calorimeter
with energy above $400$ MeV is required to satisfy $\eta^{max} < 3.2$.
The activity in the PRT and the FMD is required 
not to exceed that typical of noise
levels as obtained from randomly triggered events. 
These requirements ensure that the analysed sample is dominated by elastically
scattered protons at small $|t|$, with a small admixture of events with 
leading neutrons and low $M_{Y}$ baryon excitations, collectively referred
to here as `proton dissociation' contributions.
 
The diffractive kinematics
are reconstructed using
\begin{eqnarray}
x_{\pom} = {{\Sigma_{HFS} (E_{i} + P_{z,i})} \over {2 E_{p}^{beam}}}
\hspace*{1.3cm}
M_{X} = \sqrt{s \, y \, x_{\pom}} \ ,
\label{DeltaEtaJetsdet}
\end{eqnarray}
where $E_{p}^{beam}$ is the proton beam energy. 
A cut on $x_{\pom}$ is applied to ensure good
containment of the system $X$ and to suppress sub-leading exchange
contributions. A hadron level estimator for 
the momentum fraction $z_{\pom}$ is obtained using
\begin{eqnarray}
z_{\pom}^{\rm jets} & = & {{\Sigma_{jets} (E_{i} + P_{z,i})} 
\over {\Sigma_{HFS} (E_{i} + P_{z,i})}} \ .
\end{eqnarray}

The kinematic range in which the diffractive 
dijet measurement is performed is specified in table~\ref{tab:kinrange}.
The inclusive measurement phase space is defined by 
these conditions, with the requirements relaxed on the diffractive variables
$x_{\pom}$, $z^{\rm jets}_{\pom}$, $M_{Y}$ and $t$.
Except where measurements are made explicitly as a function of
$z_{\pom}^{\rm jets}$,  
the $z_{\pom}^{\rm jets} > 0.8$ 
region is excluded from the diffractive analysis.
This improves the reliability of the comparison between data and 
theoretical predictions, since the
DPDF sets used are 
not valid at the largest $z_{\pom}$ values.
After applying all selection criteria, about 
$3600$ out of roughly $200000$ inclusive dijet photoproduction 
events are used in the diffractive analysis.
A more detailed description of the analysis can be found
in \cite{karel:thesis}.

\renewcommand{\arraystretch}{1.35}
\begin{table}[h]
\centering
\begin{tabular}{|ccc|}
\hline
\multicolumn{3}{|c|}{Diffractive and Inclusive Measurements}     \\
\hline 
$Q^{2} <  0.01 \ {\rm GeV^2}$   & \hspace*{1cm} & $0.3 \ < y < \ 0.65$ \\ 
$E_{T}^{\rm jet1} >  5 \ {\rm GeV}$ & \hspace*{1cm} & $E_{T}^{\rm jet2} > 4 \ {\rm GeV}$ \\
$-1 < \eta^{\rm jet1} < 2$          & \hspace*{1cm} & $-1 < \eta^{\rm jet2} < 2$ \\
\hline
\multicolumn{3}{|c|}{Diffractive Measurement}  \\ \hline
$x_{\pom}  <  0.03$             & \hspace*{1cm} & $z^{\rm jets}_{\pom}  <  0.8$ \\
$M_{Y}  <  1.6 \ {\rm GeV}$     & \hspace*{1cm} & $\left| t \right| < 1 \ {\rm GeV^2}$ \\ \hline 
\end{tabular}
\caption{Kinematic ranges of the diffractive and inclusive measurements.}
\label{tab:kinrange}
\end{table}
\renewcommand{\arraystretch}{1.}

\subsection{Cross Section Measurement}
\label{sec:xsec}

The diffractive 
differential cross section is measured in each bin $i$ of a variable $x$ 
using the formula
\begin{eqnarray}
\left( {{{\rm d} \sigma} \over {{\rm d} x}} \right)_{i} & = & {{N^{data}_{i}/\varepsilon^{trig}_{i} - N^{MC,bgd}_{i}} \over {A_{i}~\Delta^{x}_{i}}~{\cal L}}\cdot~\frac{1}{C^{pdiss}} \ .
\label{xsecformula}
\end{eqnarray}
Here, $N^{data}_{i}$ is the raw number of reconstructed events passing
the selection criteria listed in section~\ref{sec:selection} and 
$\varepsilon^{trig}_{i}$ is the trigger efficiency, obtained by reference
to an independently triggered sample
and parameterised as a function of the multiplicity of charged particle tracks.
The trigger efficiency averaged over the full measurement range
is $0.86$. 
The non-diffractive background contribution 
obtained from the PYTHIA MC simulation is denoted
$N^{MC,bgd}_{i}$ and does not extend beyond the 
few percent level for any of
the measured data points. 
The factor $A_{i}$ corrects the measurement for detector 
effects, including migrations between bins, 
to the level of stable hadrons. It is calculated from the RAPGAP MC and has 
an average value of $0.31$ for the diffractive 
analysis, most of the losses being due to the limited electron tagger
acceptance of $0.40$ integrated over the measured $y$ 
range.
The bin width is denoted $\Delta^{x}_{i}$, ${\cal L}$ is the luminosity of the 
data sample and $C^{pdiss} = 0.94 \pm 0.07$, 
evaluated using the DIFFVM MC, corrects the measurement to 
the chosen range of $M_Y$ and $t$ (table~\ref{tab:kinrange}).
In the inclusive analysis the cross section is obtained analogously 
to equation~\ref{xsecformula} except for the $N^{MC,bgd}$ 
and $C^{pdiss}$ terms, which are not relevant.

\subsection{Systematic Uncertainties}
\label{sec:systs}

Uncertainties are evaluated for all significant sources of possible
systematic bias. These sources are summarised for the 
diffractive analysis below, together with
their corresponding influences 
on the total diffractive cross section.

\vspace*{-0.5cm}
\paragraph{Energy Scale:}  
The energy scale of the HFS 
measurement is tested using the momentum balance constraint between the 
precisely reconstructed positron and
the HFS in neutral current DIS events. Dedicated data and MC samples 
are analysed and found to agree to better than $3\%$. 
The effect of a relative $3\%$ change in the energy of 
the HFS 
between the data and the MC is a $9.6 \%$ shift to the 
total diffractive dijet cross section. This arises mainly from changes
in the migration corrections across the minimum 
$E_T^{\rm jet}$ values 
and the maximum $x_{\pom}$ value of the measurement. 
The energy scale uncertainties are thus highly 
correlated between the bins of the differential cross section measurements.

\vspace*{-0.5cm}
\paragraph{Large Rapidity Gap Selection:}
A fraction of the events in the kinematic range of the analysis 
(table~\ref{tab:kinrange}) give rise to 
hadronic activity in the forward detectors or
at pseudorapidities beyond those allowed by the
$\eta^{max}$ cut in the LAr calorimeter. Corrections for
this inefficiency of the
large rapidity gap selection are made using the RAPGAP MC simulation.
The uncertainties in the correction factors are assessed through a 
study of forward energy flow in a sample of
dijet photoproduction events with leading protons tagged in 
the H1 Forward Proton Spectrometer \cite{H1:FPS}. 
RAPGAP is found to describe these migrations to 
within 10\% \cite{schenk,gp:jet}, which translates into a
$2.9 \%$ uncertainty on the measured total cross section
and uncertainties which are correlated
between bins of the differential distributions. 

\vspace*{-0.5cm}
\paragraph{Proton Dissociation:}
The model dependence 
uncertainty on the proton dissociation correction factor
($C^{pdiss}$ in equation~\ref{xsecformula})
is obtained by varying the elastic and proton
dissociation cross sections and the 
proton dissociation $M_{Y}$ and $t$ dependences in the DIFFVM MC samples, 
following \cite{H1:LRG}. 
The largest effect arises from 
varying the ratio of the 
proton-elastic to the proton-dissociative cross sections between $0.5$ and 
$2$. The resulting uncertainty on the 
measured cross section
is $7 \%$.

\vspace*{-0.5cm}
\paragraph{Model Dependence:}
The influence of the model assumptions on the acceptance and bin migration 
corrections ($A_i$ in equation~\ref{xsecformula}),
is determined in the diffractive analysis 
by varying the kinematic distributions in the
RAPGAP simulation within the limits allowed by maintaining an acceptable
description of the uncorrected data. 
The following variations
are implemented by reweighting each MC event according to the value of 
generator level kinematic variables, 
leading to the quoted systematic uncertainties on the total
cross section. 
       \begin{itemize}
       \item The $x_{\pom}$ distribution is reweighted by 
$x_{\pom}^{\pm 0.2}$, leading to a $6.0 \%$ uncertainty.
       \item The $z_{\pom}^{\rm jets}$ distribution  
is reweighted by
$z_{\pom}^{{\rm jets} \ \pm 0.3}$, leading to a $4.8 \%$ uncertainty.
       \item The $x_{\gamma}^{\rm jets}$ distribution  
is reweighted by 
$x_{\gamma}^{{\rm jets} \ \pm 0.3}$, leading to a $0.6 \%$ uncertainty.
       \item The $E_{T}^{\rm jet1}$ distribution  
is reweighted by $E_{T}^{\rm jet1 \ \pm 0.4}$, leading to a $0.8 \%$ uncertainty.
       \item The $t$ distribution is reweighted by $e^{\pm 2 t}$, 
leading to a $4.4 \%$ uncertainty.
       \item The $y$ distribution is reweighted by
$y^{\pm 0.3}$, leading to a $0.2 \%$ uncertainty.
       \end{itemize}

\vspace*{-0.5cm}
\paragraph{Electron Tagger Acceptance:} 
A dedicated 
procedure external to this analysis is
used to obtain the electron tagger acceptance~\cite{etag:acc}.
The integrated acceptance over the full 
$y$ range is known to $5 \%$, which affects the cross section normalisation.

\vspace*{-0.5cm}
\paragraph{Trigger Efficiency:} 
The procedure for parameterising the trigger efficiency
($\varepsilon^{trig}_{i}$ in equation~\ref{xsecformula})
leads to a $5\%$ uncertainty.
This covers the observed deviations of the parameterisation from the
measured efficiencies 
as a function of all variables relevant to the analysis. 
This uncertainty is treated as being uncorrelated between data points.

\vspace*{-0.5cm}
\paragraph{Luminosity:} 
The measurement of the 
integrated luminosity has an uncertainty of $1.5 \%$. This translates
directly into a $1.5 \%$ normalisation uncertainty on the measured
cross sections.

\vspace*{-0.5cm}
\paragraph{Non-diffractive background:} 
A $50 \%$ normalisation variation is applied to the 
non-diffractive background contribution
given by the PYTHIA MC model ($N^{MC,bgd}$ in equation~\ref{xsecformula}). 
The effect of this change is correlated between the data points and leads 
to a $1 \%$ uncertainty on the total cross section.

\vspace*{-0.5cm}
\paragraph{Forward Detector Noise:} 
Fluctuations in the FMD noise, leading to losses in the large rapidity
gap event selection, are evaluated for each  
run using randomly triggered events.  
The standard deviation in the run-by-run distribution of the correction
factors is used to derive a $0.5 \%$ normalisation uncertainty on the 
measured cross sections. Noise in the PRT detector is negligible.    

A similar procedure is followed to evaluate the systematic uncertainties in
the inclusive dijet analysis.
The uncertainties associated with the large rapidity gap selection 
and the model dependence are no longer relevant. 
Instead, comparisons between the PYTHIA and HERWIG MCs are used 
determine a $2 \%$ 
model dependence uncertainty on the acceptance correction 
when integrated over the full phase space
studied.
The inclusive cross section systematics are dominated
by a contribution at the $10 \%$ level from 
the HFS energy scale uncertainty. However, when forming 
the ratio of diffractive to inclusive cross sections, 
this error source cancels to good approximation, the residual uncertainty
being less than $1\%$. 
The largest remaining contribution to the systematic
uncertainty on the cross section ratio
arises from the model dependence.

The total systematic uncertainty on each data point is formed by adding
the individual contributions in quadrature. 
In the figures and tables that follow, 
the systematic uncertainties are separated into two categories:
those which are uncorrelated between data points
(the model dependence and trigger efficiency) and those which
lead to correlations between data points (all other sources).

\section{Results}
\label{sec:results}

\subsection{Diffractive Dijet Cross Sections}
\label{sec:difdata}

Cross sections are measured integrated over the full kinematic range
specified in table~\ref{tab:kinrange} and also 
single- and double-differentially 
as a function of a variety of variables which are sensitive to the 
overall event structure, the hard subprocess and the presence of remnants 
of the virtual photon and the 
diffractive exchange.
The measured differential cross sections, which correspond to
averages over the specified measurement intervals, are given numerically in 
tables~\ref{table1D} and~\ref{table2D}, where the experimental 
uncertainties and hadronisation corrections applied to the NLO
calculations are also listed.
Tables~\ref{table1Dratios} and~\ref{table2Dratios} contain
the ratios of the measurements to the 
NLO calculations, obtained using the
FR framework (section~\ref{nlo}) and the H1 2006 Fit B 
DPDFs (referred to in the following as FR Fit B).

\subsubsection{Integrated Cross Section}
\label{integrated}

The total diffractive dijet positron-proton cross section 
integrated over
the full measured kinematic range is
\begin{equation}
\sigma^{tot}_{data} =  
295 \pm 6~(\mbox{stat.}) \pm 58~(\mbox{syst.})~~\mbox{pb} 
\ .
\end{equation}
The ratio of this result to the corresponding 
FR Fit B NLO prediction is
\begin{equation}
\sigma^{tot}_{data} / \sigma^{tot}_{NLO} =  0.58 \pm 0.01~(\mbox{stat.}) \pm 0.12~(\mbox{syst.})  \pm 0.14~(\mbox{scale}) \pm 0.09~(\mbox{DPDF}) \ ,
\label{sigmatot:ratio}
\end{equation}
where the statistical and systematic uncertainties 
originate from the measurement. 
The scale uncertainty corresponds to the effect of 
simultaneously varying the  
renormalisation and factorisation scales from their central values, 
$\mu_{R} = \mu_{F} = E_{T}^{\rm jet1}$, by a factor of two in either
direction. This large ($25 \%$) scale uncertainty arises
due to the relatively low $E_{T}^{\rm jet1}$ range of this 
analysis, and is the limiting factor in the comparison between data
and theory.
The DPDF uncertainty
is obtained using the method of \cite{leshouches}, by propagating
the eigenvector decomposition of the fit uncertainties. 
If the H1 2006 Fit B DPDFs are replaced by the H1 2007 Fit Jets 
DPDFs, the result is 
$\sigma^{tot}_{data} / \sigma^{tot}_{NLO} = 0.64$, 
which is inside the quoted DPDF uncertainty. Using ZEUS DPDF SJ,
a compatible result of $\sigma^{tot}_{data} / \sigma^{tot}_{NLO} = 0.70$ 
is obtained.

Adding all uncertainties in quadrature, the ratio result 
in equation~\ref{sigmatot:ratio} implies at the
$2 \sigma$ level that 
the NLO QCD calculation, neglecting any gap destruction effects, 
yields a larger diffractive dijet photoproduction
cross section than that measured. It 
confirms the result of a previous H1 analysis
in a very similar kinematic range \cite{gp:jet} and is 
broadly as expected from theoretical calculations of rapidity 
gap survival probabilities \cite{kkmr,Klasen0.34}. 

\subsubsection{Single-Differential Cross Sections}
\label{Single-differential cross sections}

\begin{figure}[h]
\begin{center}
\epsfig{file=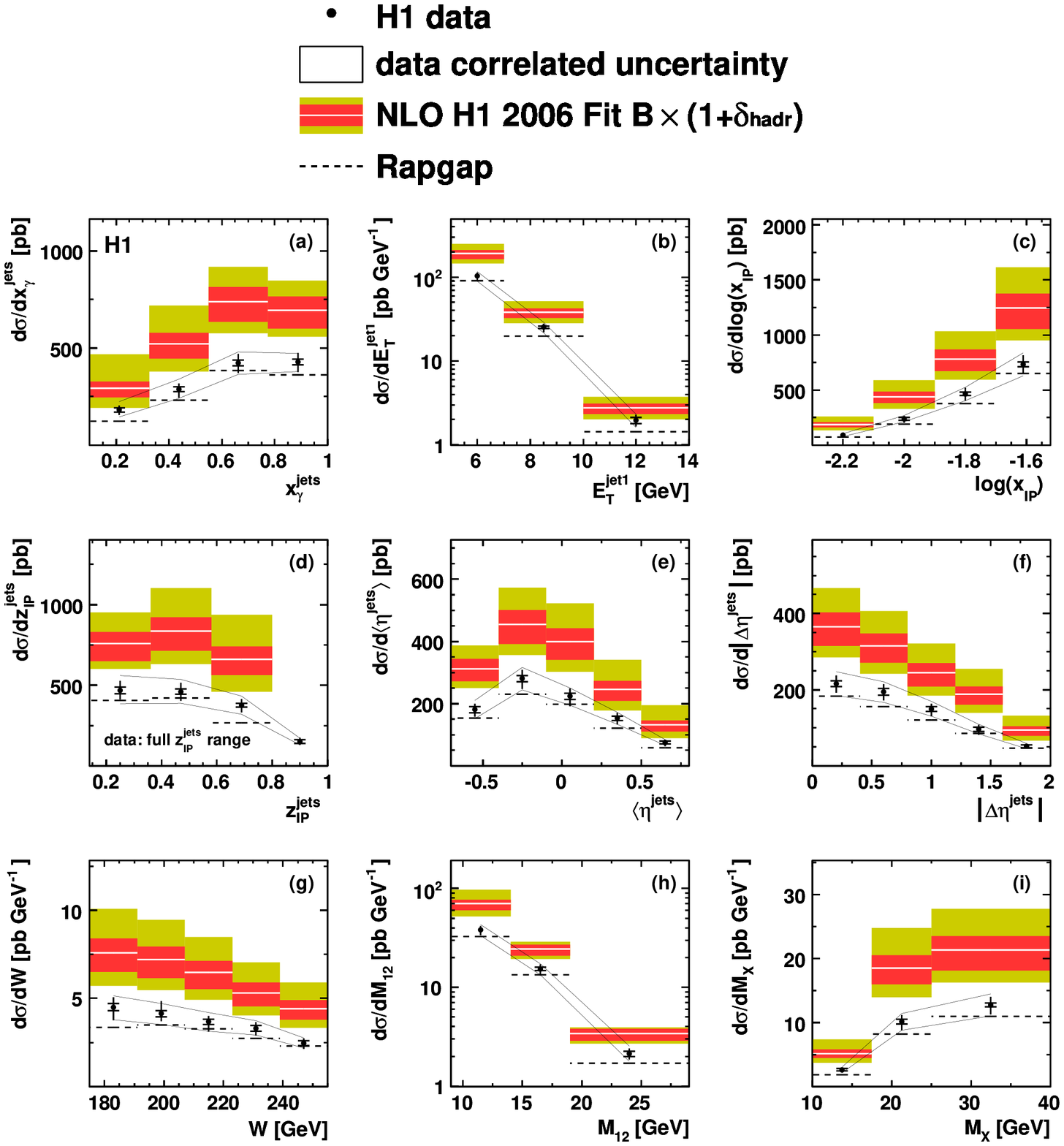 ,width=15.5cm}
\vspace*{-0.3cm}
\caption{Diffractive dijet photoproduction cross sections differential 
in (a) $x_{\gamma}^{\rm jets}$, (b) $E_{T}^{\rm jet1}$, 
(c) $\log x_{\pom}$, (d) $z_{\pom}^{\rm jets}$, 
(e) $\left<\eta^{\rm jets}\right>$, (f) $\left|\Delta\eta^{\rm jets}\right|$, 
(g) $W$, (h) $M_{12}$ and (i) $M_X$. The data points are shown with 
inner error bars corresponding to statistical uncertainties and 
outer error bars representing statistical and uncorrelated 
systematic uncertainties 
added in quadrature. The correlated 
systematic errors are indicated by the open bands between the 
two solid black lines. 
The white lines show NLO QCD calculations obtained using the FR 
framework \cite{gp:jet,FR,seb:thesis} and the H1 2006 Fit B DPDFs,
corrected for hadronisation effects. The dark bands around the
theoretical predictions indicate the result of propagating the  
uncertainties on the Fit B DPDFs to the NLO calculation. 
The light bands show 
this DPDF uncertainty added in quadrature with 
the effect on the calculation of varying $\mu_R$ and $\mu_F$ by factors of 
0.5 and 2.0. In all figures, the predictions  
of the RAPGAP MC model are also shown.}
\label{fig:sigma single}
\end{center}
\end{figure}

\begin{figure}[phh]
\begin{center}
\epsfig{file=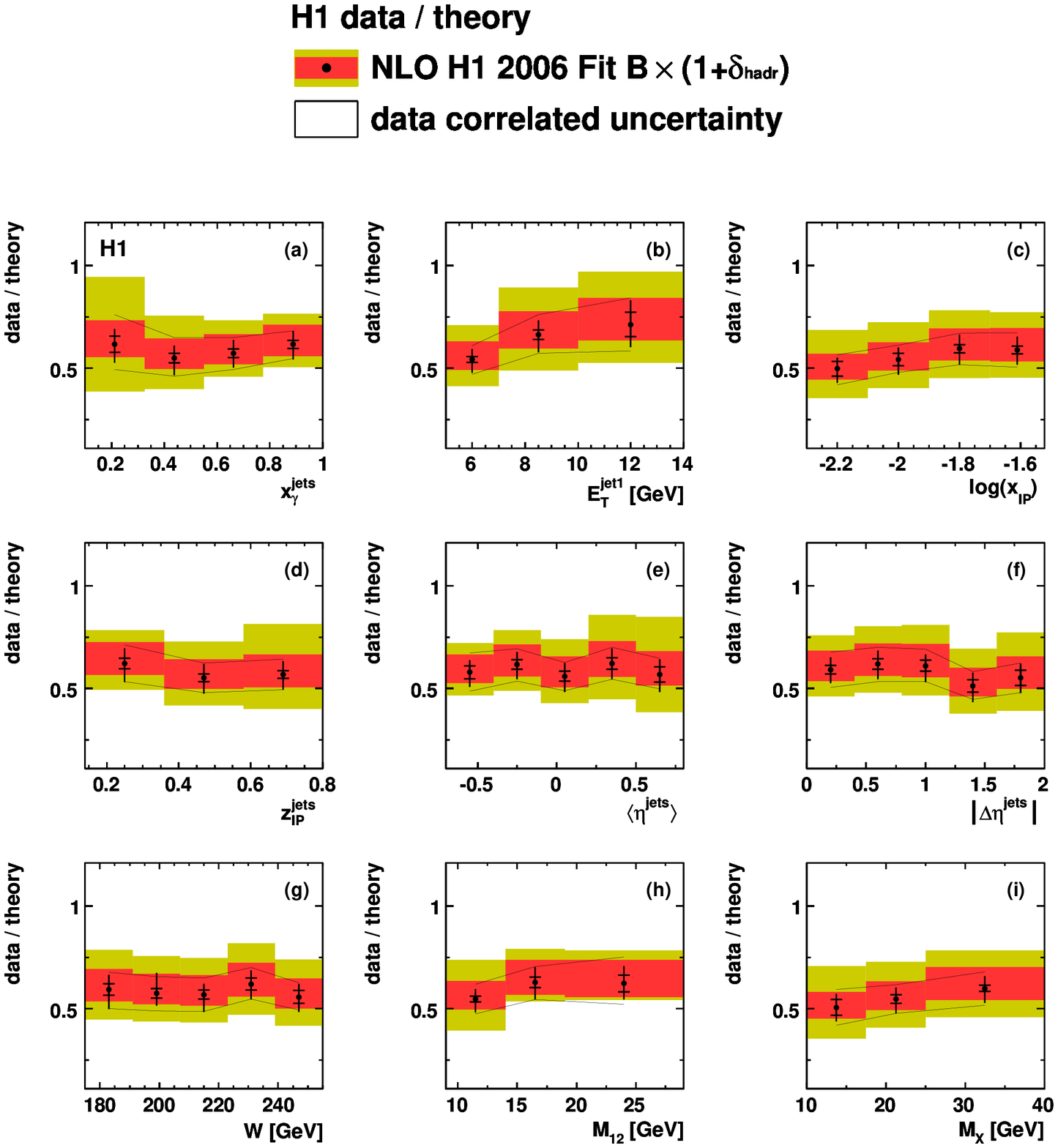 ,width=15.5cm}
\caption{Ratios of the single-differential cross sections 
to the corresponding
NLO QCD predictions
based on the FR framework and the H1 
2006 Fit B DPDF set, corrected for hadronisation effects. 
See the caption of figure~\ref{fig:sigma single} for further details.
}
\label{fig:sigma single ratio}
\end{center}
\end{figure}

Figure~\ref{fig:sigma single} shows the diffractive dijet
cross section measured single-differentially 
in $x_{\gamma}^{\rm jets}$, $E_{T}^{\rm jet1}$, $\log x_{\pom}$, 
$z_{\pom}^{\rm jets}$, $\left<\eta^{\rm jets}\right>$, 
$\left|\Delta\eta^{\rm jets}\right|$, 
$W$, $M_{12}$ and $M_X$, in the phase space defined in 
table~\ref{tab:kinrange}. In figure~\ref{fig:sigma single}(d),
the $z_{\pom}^{\rm jets} < 0.8$ requirement is relaxed and
the cross section measured for the
region $z_{\pom}^{\rm jets} > 0.8$ is shown 
without theoretical
comparisons, since the DPDFs are not 
defined (see section~\ref{sec:selection}).
To allow a more detailed shape comparison 
between the data and the predictions, 
ratios of the measured differential cross sections to the FR Fit B 
calculations are plotted in figure~\ref{fig:sigma single ratio}.   
These ratios may be taken as measurements
of the dependence of the 
rapidity gap survival probability on the
kinematic variables. 

Figures~\ref{fig:sigma single} and~\ref{fig:sigma single ratio} show that
the suppression by around a factor of $0.6$ of the data with respect to the  
FR Fit B NLO calculations has 
at most a weak dependence 
on the kinematic variables. Notably, within the uncertainties there is 
no dependence on $x_{\gamma}^{\rm jets}$ 
(figure~\ref{fig:sigma single ratio}(a)),
in contrast to theoretical predictions for the rapidity
gap survival probability \cite{kkmr,Klasen0.34}. The largest dependence
of the central values of the measured ratios
on any of the variables appears in the cross section
differential in $E_{T}^{\rm jet1}$ (figure~\ref{fig:sigma single ratio}(b)).
Although not well established by the current data, 
this dependence is compatible with
previous data \cite{gp:jet,ZEUSphp,karel:DIS08}. 
The $E_{T}^{\rm jet1}$ dependence is 
investigated further in section~\ref{double:diff}.

The measured cross sections in figure~\ref{fig:sigma single}
are also compared with a prediction obtained 
using the RAPGAP MC generator (section~\ref{sec:MC}), which does not
contain any model of rapidity gap destruction. 
The shapes of the 
measured cross sections are well described and the 
normalisation is only slightly lower than that of the data. 
However the scale uncertainty in this model is rather large
and the same model 
undershoots diffractive dijet measurements in 
DIS \cite{gp:jet,Matthias}, 
where factorisation
is expected to hold. 

In \cite{ZEUSphp}, the ZEUS collaboration presented an analysis of diffractive
dijet photoproduction data with $E_{T}^{\rm jet1} > 7.5 \ {\rm GeV}$, which 
is most readily compared with the 
second and third $E_{T}^{\rm jet1}$ intervals
in figures~\ref{fig:sigma single}(b) 
and~\ref{fig:sigma single ratio}(b). 
However, even for $E_{T}^{\rm jet1} > 7.5 \ {\rm GeV}$,
a direct comparison between
H1 and ZEUS data is not possible,
since the ZEUS analysis covers a wider $y$ range and cuts
on the second jet at an
$E_{T}^{\rm jet2}$ value of $6.5 \ {\rm GeV}$, larger than the
value used here. 
An indirect comparison can be made on the basis of ratios of the data to
 NLO theoretical calculations using the H1 Fit B DPDFs.
ZEUS obtains a result of around $0.9$ for this ratio, which is compatible 
with the result for $E_{T}^{\rm jet1} > 7.5 \ {\rm GeV}$ obtained here, within the
large combined uncertainties.\footnote{In \cite{zeus:f2d,newman:ruspa} a 
13\% difference
between H1 and ZEUS inclusive DDIS 
data is identified. This is within the 
combined normalisation uncertainties of the two experiments, 
which are largely
due to proton dissociation. If the dijet photoproduction cross sections
in the two experiments are normalised to the inclusive DDIS 
data \cite{H1:LRG,zeus:f2d}, the remaining differences in the common 
$E_{T}^{\rm jet1}$ range are well within the experimental uncertainties alone.}

\begin{figure}[phh]
\begin{center}
\epsfig{file=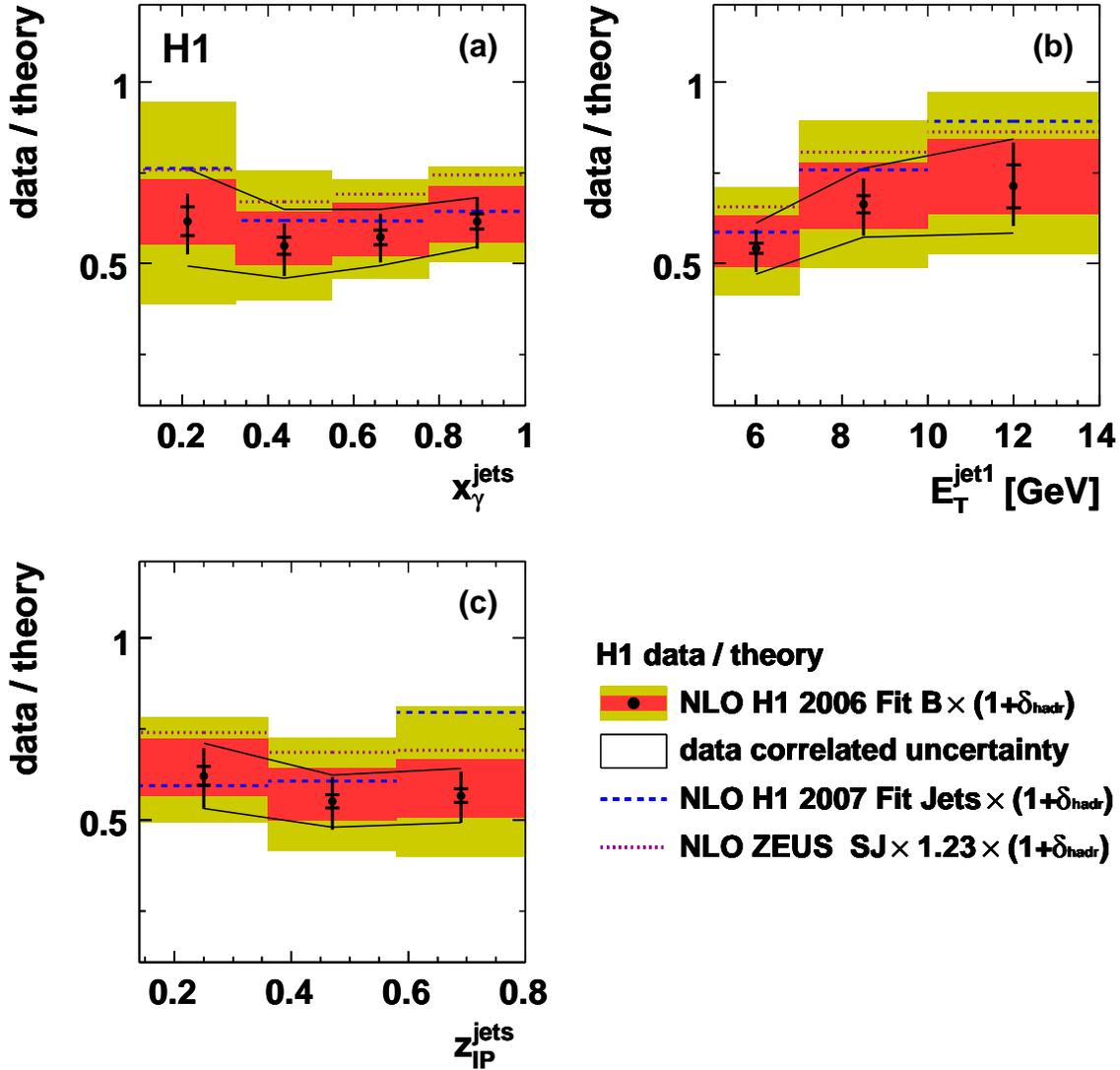 ,width=15.5cm}
\caption{Ratios of the single-differential cross sections 
to the corresponding
NLO QCD predictions
based on the FR framework, corrected for hadronisation effects,
for a subset of variables. Results obtained with H1 2006 Fit B set
of diffractive parton densities
are compared with those from H1 2007 DPDF Fit Jets and from the
ZEUS DPDF SJ fit. The latter is scaled by a factor 
of $1.23$ \cite{H1:FPS} to convert its coverage from elastic protons only
($Y = p$)
to $M_Y < 1.6 \ {\rm GeV}$. 
See the caption of figure~\ref{fig:sigma single} for further details.}
\label{fig:ratiodpdfs}
\end{center}
\end{figure}

As discussed in detail in \cite{H1:LRG,Matthias}, the error
bands on the DPDFs extracted from inclusive diffraction alone
do not include uncertainties due to parton parameterisation choices 
and thus do not reflect the full uncertainties, particularly in the large 
$z_{\pom}^{\rm jets}$ region. 
To give a complementary indication
of the possible range
of variation, comparisons between the ratios obtained
with the H1 2006 Fit B DPDFs, the H1 2007 Fit Jets DPDFs and 
ZEUS DPDF SJ fit are shown for a subset of variables ($x_{\gamma}^{\rm jets}$, 
$E_{T}^{\rm jet1}$ and $z_{\pom}^{\rm jets}$)
in figure~\ref{fig:ratiodpdfs}. The ZEUS DPDFs lead to ratios which 
are uniformly $10-15\%$ larger than those obtained with H1 2006 Fit B,
with no strong dependence on any of the kinematic variables.
The deviation of the H1 2007 Fit Jets result from the
H1 2006 Fit B result extends beyond the DPDF error
band for $z_{\pom}^{\rm jets} > 0.6$, which is correlated with a 
somewhat stronger dependence of the ratio of 
data to theory on $E_{T}^{\rm jet1}$
and a slightly different shape at low $x_{\gamma}^{\rm jets}$.

\begin{figure}[phh]
\begin{center}
\epsfig{file=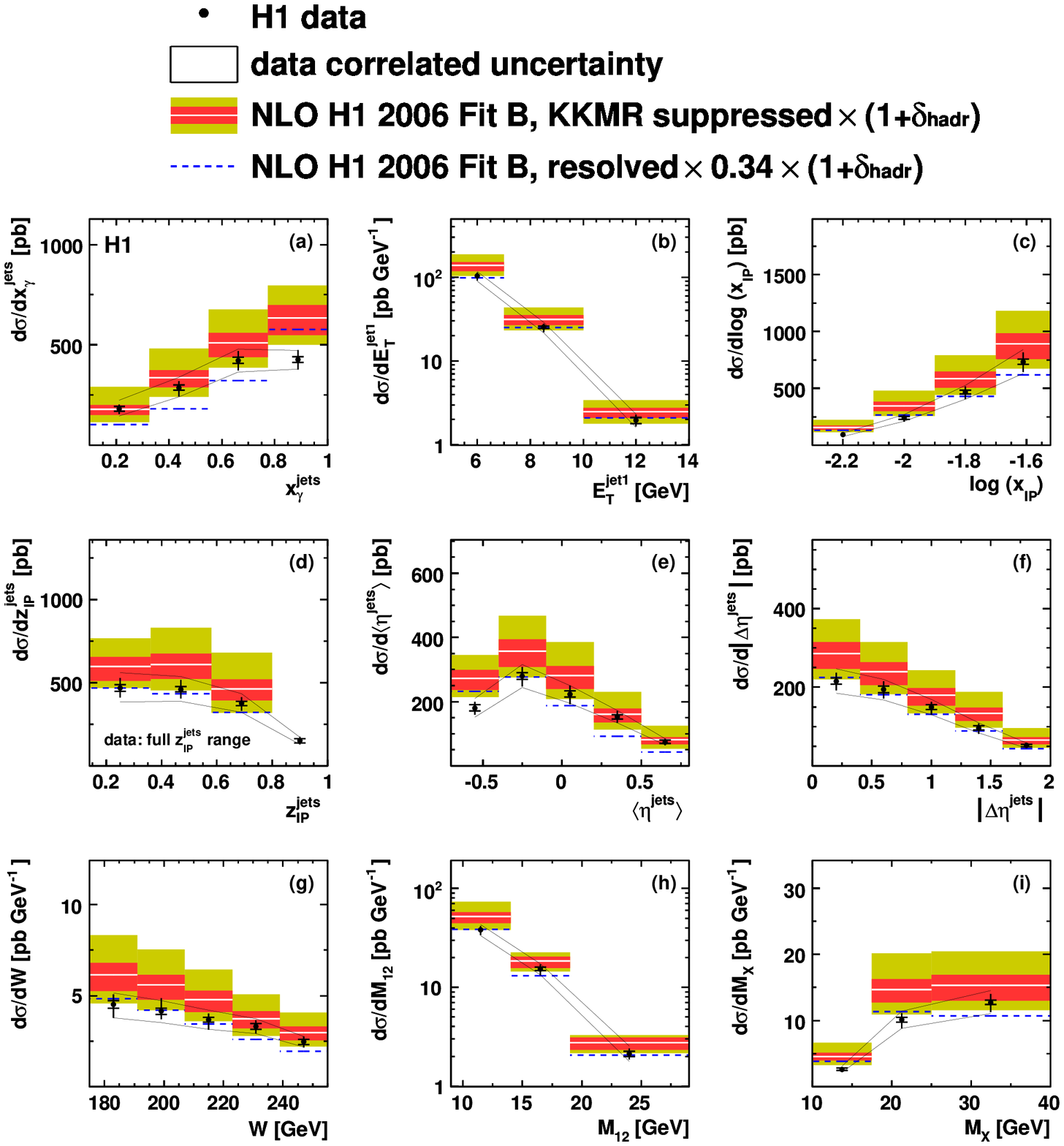 ,width=15.5cm}
\caption{Single-differential diffractive dijet photoproduction
cross sections as in figure~\ref{fig:sigma single}.
The FR theoretical prediction for resolved photons 
is modified by applying the scale factors from the KKMR 
model 
for point-like interactions (`KKMR suppressed') \cite{kkmr:new} or
for hadron-like interactions (`resolved $\times 0.34$') \cite{kkmr}. 
The direct photon contribution is
left unchanged in both cases.
See the caption of figure~\ref{fig:sigma single} for further details.} 
\label{fig:sigma single res supp} \end{center}
\end{figure}

\begin{figure}[phh]
\begin{center}
\epsfig{file=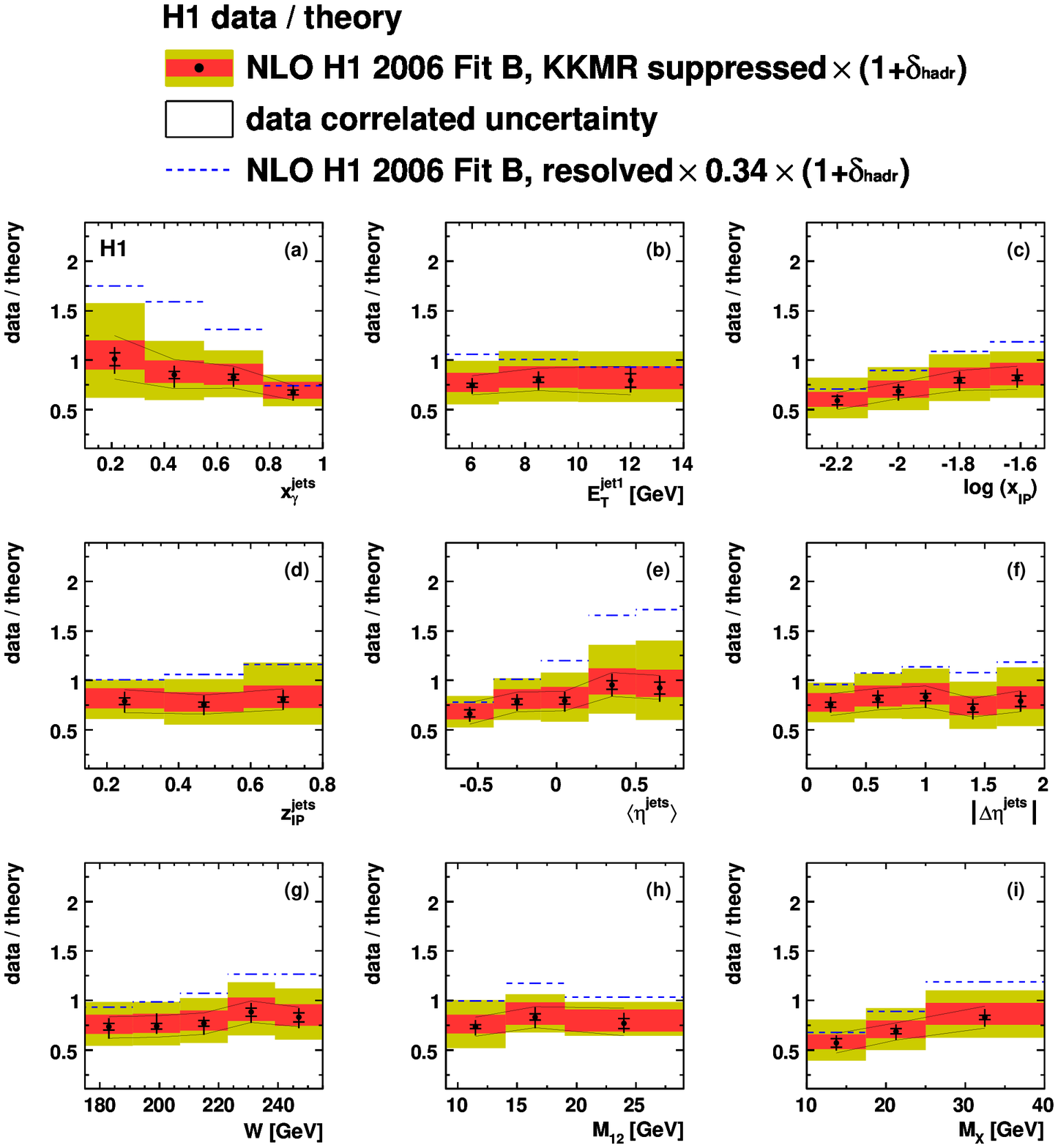 ,width=15.5cm}
\caption{Ratios of the single-differential cross sections 
to the corresponding
NLO QCD predictions
based on the FR framework and the H1 
2006 Fit B DPDF set, corrected for hadronisation effects.
The FR theoretical prediction for resolved photons 
is modified by applying the scale factors from the KKMR 
model 
for point-like interactions (`KKMR suppressed') \cite{kkmr:new} or
for hadron-like interactions (`resolved $\times 0.34$') \cite{kkmr}. 
The direct photon contribution is
left unchanged in both cases.
See the caption of figure~\ref{fig:sigma single} for further details.}
\label{fig:sigma single ratio res supp} \end{center}
\end{figure}

According to the RAPGAP model,
approximately half of the cross section in the 
kinematic range studied arises from each of the direct and 
resolved photon-induced contributions.  
The decomposition of photoproduction processes into 
direct and resolved
interactions is not uniquely defined beyond LO.
When modelling rapidity gap survival probabilities in the 
following, the resolved photon contribution is
defined to correspond exactly to that which is calculated using
the photon structure function.\footnote{In \cite{Klasen0.34}, 
an alternative procedure is introduced, whereby
the part of the direct
contribution which depends on the factorisation scale at the photon
vertex is also suppressed, stabilising the dependence of the
combined direct and resolved cross 
sections on this scale \cite{klasen:scale}. 
The difference between the
rapidity gap survival probabilities obtained 
using the two methods
(6\% in fits \cite{Klasen0.34} to previous H1 data \cite{gp:jet}) 
is small in comparison to other uncertainties.}
Following the calculation using an 
absorptive model of a gap survival probability of $0.34$
for the hadron-like component of resolved photoproduction \cite{kkmr}, 
previous H1 data \cite{gp:jet} 
were compared in \cite{klasen} with 
predictions in which the full
resolved photon contribution was suppressed by this factor,
the direct photon contribution being left unsuppressed.
In a later analysis \cite{Klasen0.34}, this procedure was
extended to NLO. The conclusions of these previous 
studies are 
confirmed in figures~\ref{fig:sigma single res supp} 
and~\ref{fig:sigma single ratio res supp}
through a
similar comparison of the current data with NLO calculations
in which the resolved photon contribution is globally suppressed
by a factor of 0.34.
The overall normalisation of this calculation 
is in good agreement with the data. 
However, the shapes of some of the differential distributions are
not well reproduced. In particular, there is a variation by more than a 
factor of two 
in the ratio of data to theory as a function of 
$x_{\gamma}^{\rm jets}$ (figure~\ref{fig:sigma single ratio res supp}a). 

The distinction between point-like and hadron-like resolved photon
interactions recently developed in \cite{kkmr:new} leads to a significantly 
weaker predicted suppression in the kinematic range of the current
analysis. The data are compared with this refined `KKMR'
model under the approximation of completely neglecting
hadron-like
resolved photon contributions, which, according to the authors, become 
dominant only for
$x_{\gamma}^{\rm jets} < 0.1$ \cite{kkmr:private}, beyond 
the range of the current analysis. 
The rapidity gap survival probabilities 
obtained in \cite{kkmr:new} for point-like photon interactions using the GRV 
HO photon PDFs are applied to all resolved photon interactions.
Interactions involving quarks and gluons from the photon are thus
suppressed by factors of $0.71$ and $0.53$, respectively.
The quark-initiated contribution is dominant throughout the measured
range, such that the rapidity gap survival probability in the model
is approximately $0.7$ for resolved photon interactions and $1$ for
direct photon interactions. 

Figure~\ref{fig:sigma single res supp} shows the comparison between the
measured single differential cross sections and the NLO QCD predictions, 
with the resolved photon contribution scaled according to the 
KKMR model. 
The corresponding ratios 
of data to theoretical predictions are shown
in figure~\ref{fig:sigma single ratio res supp}. 
The overall normalisation of the KKMR-based calculation
is larger than that of the data, but is compatible within
the large uncertainties.
Many of the distributions studied
are well described in shape ($E_{T}^{\rm jet1}$,  
$z_{\pom}^{\rm jets}$, $\left|\Delta\eta^{\rm jets}\right|$, 
$W$ and $M_{12}$). The data
thus agree with the prediction \cite{Klasen0.34}
that the $E_{T}^{\rm jet1}$ dependence of the data/theory ratio
flattens if the resolved photon contribution alone is suppressed. 
However, there remains a
variation in the ratio of data to the KKMR model with 
$x_{\gamma}^{\rm jets}$
and to a lesser extent with $\left<\eta^{\rm jets}\right>$, $x_{\pom}$ and $M_X$.
A comparison of figures~\ref{fig:sigma single ratio} 
and~\ref{fig:sigma single ratio res supp} shows that the 
shapes of the differential cross sections are
generally better described with a global suppression factor than with
a survival probability applied to resolved photon interactions only. 

\subsubsection{Double-Differential Cross Sections}
\label{double:diff}

To study further the dynamics of rapidity gap suppression
and their dependence on the nature of the photon interaction, cross 
sections are measured double differentially in two regions of
$x_{\gamma}^{\rm jets}$, which are
enriched with either resolved ($x_{\gamma}^{\rm jets} < 0.75$) or 
direct ($x_{\gamma}^{\rm jets} > 0.75$) 
photon processes. Using the RAPGAP MC model with the GRV-G LO photon PDFs, the 
$x_{\gamma}^{\rm jets} < 0.75$ region is estimated to contain $77\%$ resolved
photon interactions integrated over the measurement region
(table~\ref{tab:kinrange}),
with a $75\%$ direct photon contribution for
$x_{\gamma}^{\rm jets} > 0.75$. 

\begin{figure}[phh]
\begin{center}
\epsfig{file=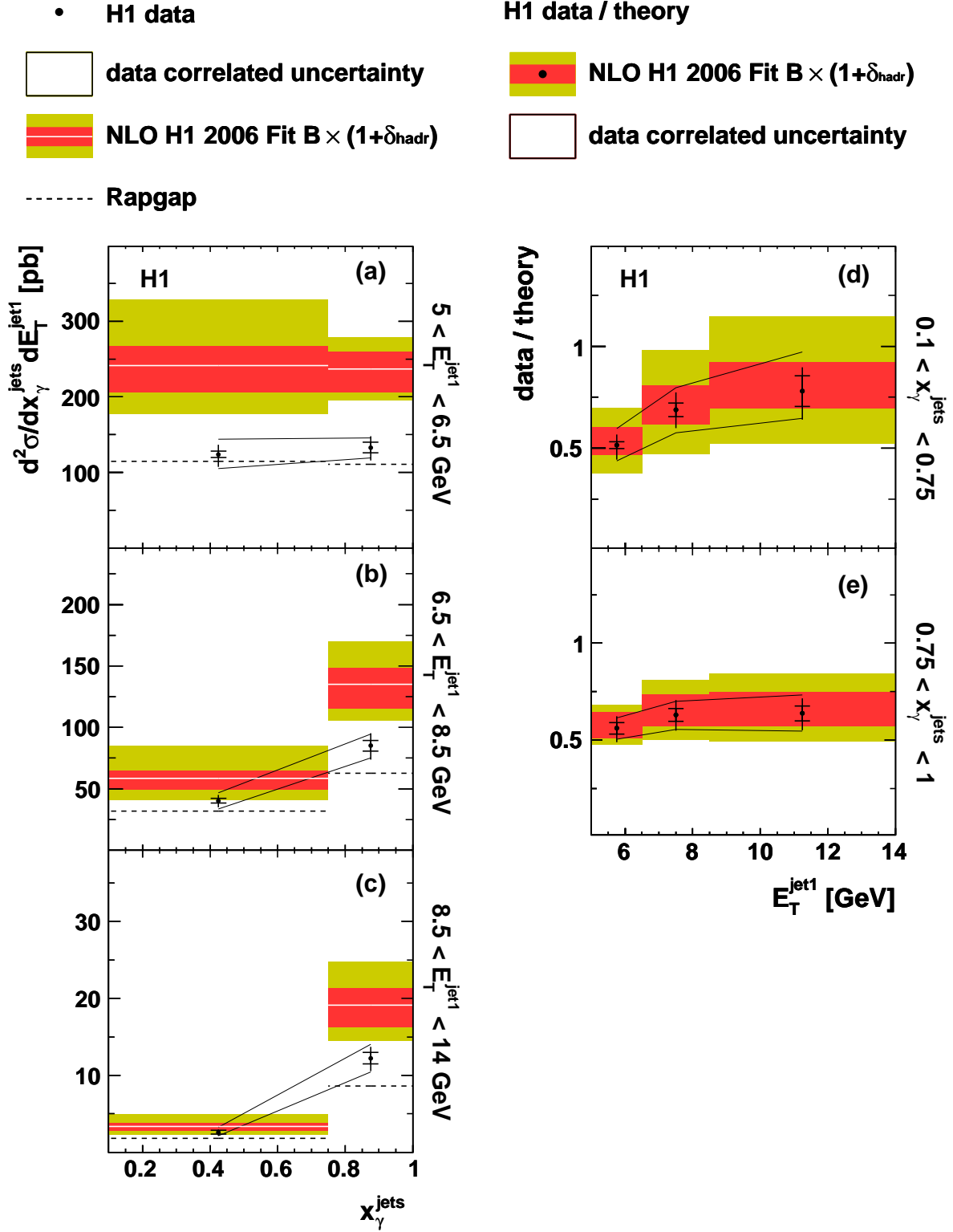 ,width=16.cm}
\caption{(a)-(c) Double-differential cross section 
${\rm d}^2 \sigma / {\rm d} E_{T}^{\rm jet1} {\rm d} x_{\gamma}^{\rm jets}$
as a function of $x_{\gamma}^{\rm jets}$, compared with NLO QCD and
RAPGAP predictions. 
(d)-(e) Ratio of the double-differential cross section to the NLO 
prediction as a function of $E_{T}^{\rm jet1}$. 
See the caption of figure~\ref{fig:sigma single} for further details.}
\label{fig:sigma et xg 1}
\end{center}
\end{figure}

In figures~\ref{fig:sigma et xg 1}(a)-(c), measurements are 
presented of the double-differential dijet cross section 
${\rm d}^2 \sigma / {\rm d} E_T^{\rm jet1} {\rm d} x_{\gamma}^{\rm jets}$
for three $E_{T}^{\rm jet1}$ ranges in the resolved and direct 
photon-enriched $x_{\gamma}^{\rm jets}$ intervals. The data are compared 
with the FR Fit B calculations and with the RAPGAP MC predictions.
Due to kinematic constraints, the resolved-enriched 
cross section at low $x_{\gamma}^{\rm jets}$ 
falls most rapidly as $E_{T}^{\rm jet1}$ increases.
There is a suggestion that this $E_{T}^{\rm jet1}$ dependence in the 
resolved-enriched region is stronger for the NLO QCD theory than for the data. 
In the direct-enriched high $x_{\gamma}^{\rm jets}$ region,
the cross section falls more slowly with $E_{T}^{\rm jet1}$ and the dependence
in the data is similar to that predicted by the NLO calculation. 
These features are illustrated further
in figures~\ref{fig:sigma et xg 1}~(d)-(e), 
where the ratios of the 
data to the NLO theory from figures~\ref{fig:sigma et xg 1}~(a)-(c) 
are presented as a function of $E_{T}^{\rm jet1}$ in the resolved and direct 
photon-enriched $x_{\gamma}^{\rm jets}$ regions, 
respectively. 

The significance of the $E_{T}^{\rm jet1}$
dependence in the resolved-enriched region 
(figure~\ref{fig:sigma et xg 1}~(d)) is evaluated through a 
$\chi^2$ test.
All uncertainties are taken into account in this procedure,
though the main contribution comes from the statistical and 
uncorrelated systematic uncertainties on the data,  
the remaining
uncertainties changing only the
normalisation of the ratio to first approximation. 
A test of
the hypothesis that there is no 
$E_{T}^{\rm jet1}$ 
dependence yields a $\chi^2$ value of 1.36, with two degrees of 
freedom, corresponding to an $E_{T}^{\rm jet1}$ 
variation at the 73\% confidence
level. 
The suppression of the data relative to the NLO prediction in the 
direct-enriched 
large $x_{\gamma}^{\rm jets}$ region is, within errors, independent 
of $E_{T}^{\rm jet1}$ (figure~\ref{fig:sigma et xg 1}~(e)).
Figures~\ref{fig:sigma et xg 1}~(d)-(e) thus indicate that any 
$E_{T}^{\rm jet1}$
dependence of the data-to-theory ratio in 
figure~\ref{fig:sigma single ratio}(b)
is driven primarily by resolved photon interactions. An
$E_{T}^{\rm jet1}$ dependence of the gap survival probability  
is predicted in the KKMR model, due to variations in 
the size of the $q \bar{q}$ dipole produced by the
point-like photon splitting, and hence in the absorptive correction. 
However the predicted effect is small
($4\%$ as $E_{T}^{\rm jet1}$ 
changes from $5 \ {\rm GeV}$ to $7.5 \ {\rm GeV}$). 
Figures~\ref{fig:sigma single ratio}(d)-(e) also indicate that 
when $E_{T}^{\rm jet1}$ becomes large, the 
suppression in the direct region may be stronger than that in the resolved
region, which is not expected in any model. 
The large uncertainties permit
statistical fluctuations in the data or small inadequacies in the theory
as possible explanations.

\begin{figure}[phh]
\begin{center}
\epsfig{file=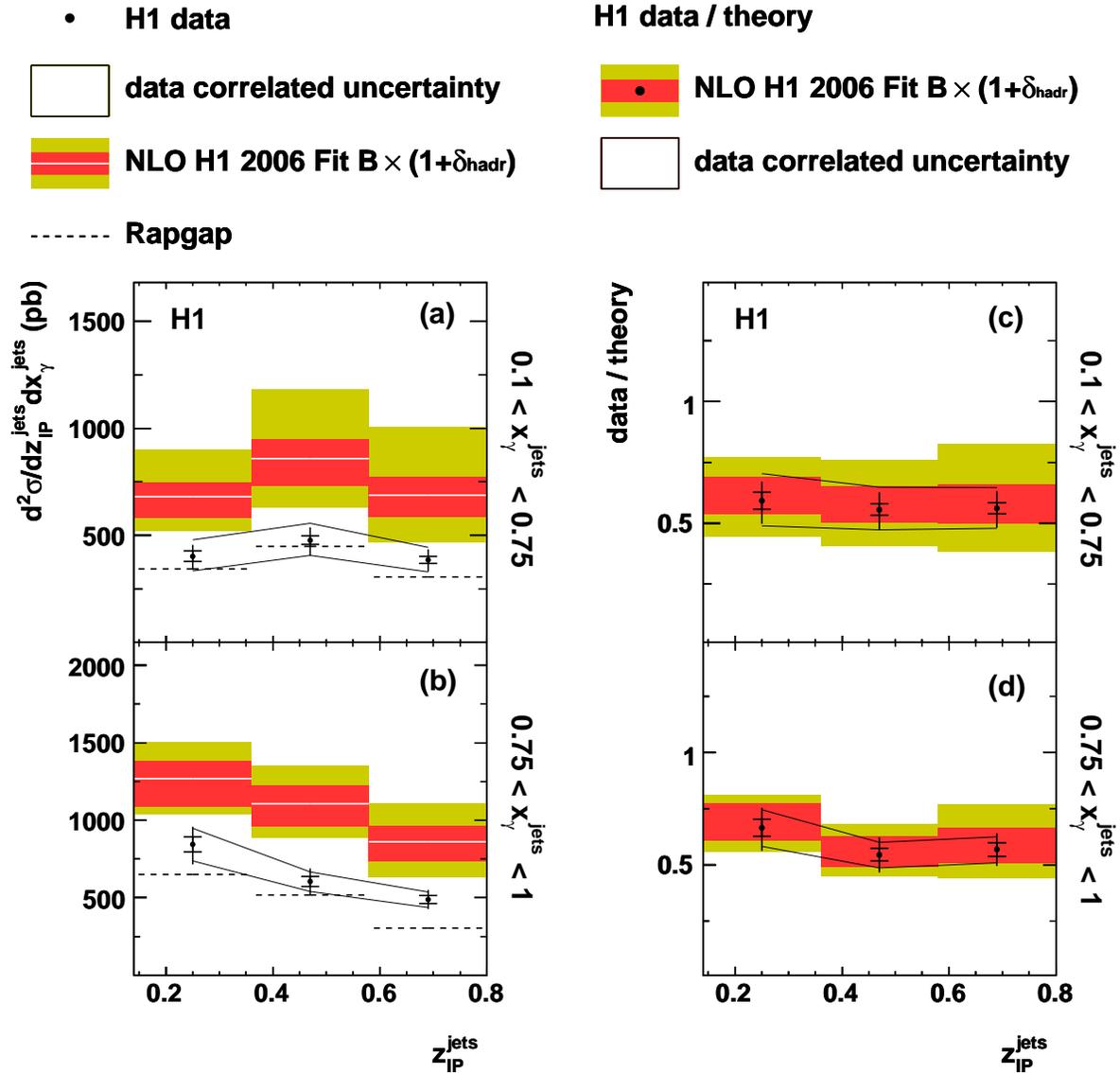 ,width=16.cm}
\caption{(a)-(b) Double-differential cross section 
${\rm d}^2 \sigma / {\rm d} z_{\pom}^{\rm jets} {\rm d} x_{\gamma}^{\rm jets}$. 
(c)-(d) Ratio of the measured double-differential cross section 
to the NLO prediction using the H1 2006 Fit B and H1 2007 Fit Jets DPDFs. 
See the caption of figure~\ref{fig:sigma single} for further details.}
\label{fig:sigma zp xg 1}
\end{center}
\end{figure}

In figure~\ref{fig:sigma zp xg 1}, the cross section 
is shown double differentially in 
$z_{\pom}^{\rm jets}$ and $x_{\gamma}^{\rm jets}$. 
The measured cross section is compared with the NLO theory as a function of 
$z_{\pom}^{\rm jets}$ 
in two bins of $x_{\gamma}^{\rm jets}$ in 
figures~\ref{fig:sigma zp xg 1}~(a)-(b) and the ratios of data to 
theoretical predictions are shown in figures~\ref{fig:sigma zp xg 1}~(c)-(d). 
The NLO calculations describe the measured shapes rather well, with no
evidence for any variation of the suppression factor between any of the 
measurement ranges. The gap survival probability 
in a region where 
there are small or no remnants of 
either the photon or the diffractive exchange
(highest $z_{\pom}^{\rm jets}$ bin in figure~\ref{fig:sigma zp xg 1}d)
is thus similar to that where both remnants are significant
(lowest $z_{\pom}^{\rm jets}$ 
bin in figure~\ref{fig:sigma zp xg 1}c). This remains the case
when the H1 2007 Fit Jets DPDFs are used in place of H1 2006 Fit B.
In both figures~\ref{fig:sigma et xg 1} and~\ref{fig:sigma zp xg 1}, 
the RAPGAP MC prediction gives a satisfactory 
description of the shapes of the double differential cross sections, 
the normalisation being slightly lower than that of the data.

\subsection{Ratios of Diffractive to Inclusive Cross Sections}

Measurements of ratios of diffractive to inclusive dijet photoproduction
cross sections have
been proposed \cite{kkmr,klasen,Klasen0.34} as a further
test of gap survival issues. 
Their potential advantages over
straight-forward diffractive measurements lie in the 
partial cancellations of some experimental systematics and of
theoretical uncertainties due to the photon structure and factorisation
and renormalisation scale choices.
The sensitivity to absorptive effects 
of diffractive-to-inclusive ratios is thus
potentially superior to that of pure diffractive cross sections.
For the ratio extraction presented here, inclusive dijet cross sections are 
measured using data 
collected in the same period as the diffractive 
sample. The experimental method and systematic error treatment
for the inclusive case is described in 
section~\ref{sec:procedure}. It is identical to the diffractive measurement
method, with the exception of the large rapidity gap requirements.

At the relatively low transverse energies studied in the present analysis,
underlying event effects have a 
large influence on 
jet cross sections
in inclusive photoproduction~\cite{MI}.
Here, the PYTHIA and HERWIG MC models are used
to correct the inclusive data for detector effects, with MI
included as described in section~\ref{sec:MChad}. 
The two models agree rather well on the corrections to be applied to the
data. The average of the results with the two models is
therefore used to calculate the corrections 
and the uncertainty. The latter is taken from the difference between the 
results with the two models and
is relatively small ($2 \%$ when integrated over the full measured range). 

\begin{figure}[phh]
\begin{center}
\epsfig{file=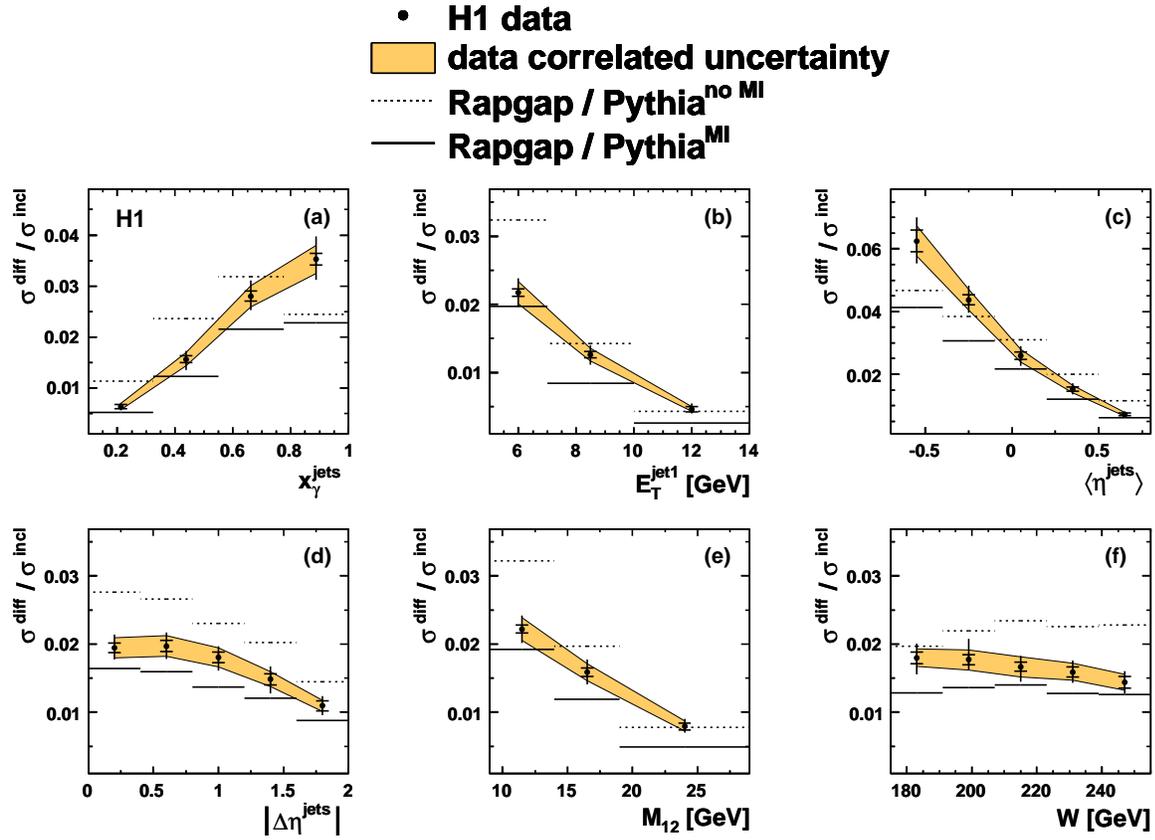 ,width=16.cm}
\caption{Ratios of diffractive to inclusive single-differential
cross sections
as a function of (a) $x_{\gamma}^{\rm jets}$, (b) $E_{T}^{\rm jet1}$, 
(c) $\left<\eta^{\rm jets}\right>$, (d) $\left|\Delta\eta^{\rm jets}\right|$, 
(e) $M_{12}$ and (f) $W$. 
The data points are shown with 
inner error bars corresponding to statistical uncertainties and 
outer error bars representing statistical and uncorrelated uncertainties 
added in quadrature. The correlated 
systematic errors are indicated by the hatched bands. 
The dashed lines 
represent the predictions from the ratio of the RAPGAP diffractive to the 
PYTHIA inclusive MC models without a multiple parton interaction model 
included in PYTHIA. The solid 
lines correspond to the same ratio of MC models with the inclusion
of multiple parton interactions in PYTHIA.}
\label{fig:R}
\end{center}
\end{figure}

The ratios 
of diffractive to inclusive single-differential dijet cross sections are 
given numerically in table~\ref{tabledifftoincl} and are 
shown in figure~\ref{fig:R} as a function of 
$x_{\gamma}^{\rm jets}$, $E_{T}^{\rm jet1}$, $\left<\eta^{\rm jets}\right>$, 
$\left|\Delta\eta^{\rm jets}\right|$, 
$M_{12}$ and $W$.
Due to the partial or complete
cancellations of some error sources
when forming the ratio, 
the correlated uncertainties are reduced compared
with those for the diffractive distributions. 

Since they give adequate descriptions of the diffractive and inclusive
data, respectively, the RAPGAP and PYTHIA MC models 
are used to assess the relative sensitivity of the 
diffractive-to-inclusive ratio to the gap
survival and MI effects. With no MI effects included in the PYTHIA 
model, the description of the inclusive data is poor and the
ratio of RAPGAP to PYTHIA exceeds the data by a factor of
around 1.5. As expected, this factor becomes smaller as $E_{T}^{\rm jet1}$
increases. However, the shape of the prediction also differs from that
of the ratio data for most of the other variables
studied, in particular $x_{\gamma}^{\rm jets}$. 

The inclusion of the PYTHIA MI model changes the
predicted inclusive cross sections, and hence the ratios, substantially.
The ratio of RAPGAP to PYTHIA then gives an improved description of the
shapes of the distributions.
The MI effects alter the 
predicted ratio by a factor of 0.5 at 
low $x_{\gamma}^{\rm jets}$, where the resolved photon remnant is most
important. As expected, there is little effect in
the direct photon-dominated large $x_{\gamma}^{\rm jets}$ region. 
The normalisation of the ratio of the models when MI are included
is smaller than that of the data. This
partially reflects the 
RAPGAP description of the diffractive data (figure~\ref{fig:sigma single})
and is partially due to an overshoot in
the PYTHIA description of the inclusive data. 

The fractional reduction in the predicted inclusive cross section
when MI are introduced in the PYTHIA model
is comparable to the magnitude of the gap survival 
suppression factor in the diffractive data (section~\ref{sec:difdata}).
The uncertainties in modelling 
the MI are large and difficult to quantify.
The precision with which gap survival issues can be unfolded from MI
complications in the ratio of diffractive to inclusive data
is correspondingly poor. Therefore no
strong conclusions can be drawn with our current understanding of MI,
despite the relatively good precision of the data.

\section{Summary}

Single and double-differential cross sections are measured for diffractive 
dijet photoproduction and are compared with 
predictions based on NLO QCD calculations 
using different sets of DPDFs. Ratios of the measured to the 
predicted differential cross sections are also studied. 

The total diffractive dijet cross section is overestimated by the NLO QCD 
theory by about a factor of two, which is consistent 
with previous H1 measurements \cite{gp:jet}. 
The shapes of the single-differential cross sections are 
well described when the H1 2006 DPDF Fit B partons are used.
A good overall description of the differential 
cross sections is obtained by applying a global suppression 
factor of $0.58 \pm 0.21$ to the NLO 
calculations.
As in similar previous analyses \cite{gp:jet,ZEUSphp,karel:DIS08}, 
there is a suggestion 
of a dependence of the rapidity gap survival probability on
$E_{T}^{\rm jet1}$, though the significance of this effect is not large. 

If only the resolved photon contribution in the calculation
is suppressed by 
a factor of $0.34$, as 
predicted for hadron-like resolved photon interactions \cite{kkmr}, 
the overall normalisation of the NLO QCD prediction agrees well with the 
data. However, the description of the $x_\gamma^{\rm jets}$ distribution,
which best distinguishes direct  
from resolved photon interactions, becomes poor. 
If rapidity gap survival probabilities expected for point-like 
resolved photons are applied instead \cite{kkmr:new}, the overall normalisation
is acceptable and the $E_{T}^{\rm jet1}$ dependence of the data is better
described. However, the description of the 
$x_\gamma^{\rm jets}$ dependence remains problematic.

The analysis of the double-differential cross section  
${\rm d}^2 \sigma / {\rm d} E_T^{\rm jet1} {\rm d} x_{\gamma}^{\rm jets}$ 
indicates that the $E_{T}^{\rm jet1}$ dependence of the data/theory ratio
originates from 
the resolved photon-enriched region of $x_{\gamma}^{\rm jets}$.
However, the data are also consistent with no dependence on $E_{T}^{\rm jet1}$
for either of the $x_{\gamma}^{\rm jets}$ regions studied. 
The ratio of the 
data to the NLO theory for the double-differential cross section 
${\rm d}^2 \sigma / {\rm d} z_{\pom}^{\rm jets} {\rm d} x_{\gamma}^{\rm jets}$
is constant within
errors throughout the region studied, indicating that the gap survival
probability is insensitive to the presence or nature of remnants of either
the photon or the diffractive exchange.

Measurements of the ratio 
of diffractive to inclusive single-differential 
cross sections are presented as a function of several variables. 
The influence of multiple interaction effects in the inclusive data 
is large in
the kinematic range studied here. 
The large uncertainties in modelling 
these multiple interactions preclude strong conclusions
about rapidity gap survival on the basis of these data,
although a reasonable description of the ratios
can be obtained with suitably tuned 
Monte Carlo models.

\section*{Acknowledgements}

We are grateful to the HERA machine group whose outstanding
efforts have made this experiment possible. 
We thank the engineers and technicians for their work in constructing and
maintaining the H1 detector, our funding agencies for 
financial support, the
DESY technical staff for continual assistance
and the DESY directorate for support and for the
hospitality which they extend to the non-DESY 
members of the collaboration.
We express our gratitude to A.~Kaidalov, V.~Khoze, M.~Klasen, G.~Kramer, 
A.~Martin and M.~Ryskin for many helpful discussions.

\newpage

%
%
\begingroup
\everymath{\displaystyle}
\footnotesize
\begin{table}[htb!]
\begin{center}
{\footnotesize
\begin{tabular}{|ccc|c|c|c|c|c|}
\hline
&&&&&&& \\[-10pt]
\multicolumn{3}{|c|}{$x^{\rm jets}_{\gamma}$} & $\mbox{d}\sigma/\mbox{d}x^{\rm jets}_{\gamma}\,\,[\mbox{pb}]$ & $\delta_{\mbox{\tiny stat}}$ & $\delta_{\mbox{\tiny uncorr}}$ & $\delta_{\mbox{\tiny corr}}$ & $1+\delta_{\mbox{\tiny hadr}}$ \\[0.1cm]
\hline
${\scriptstyle 0.1}$&\!\!\!\!${\scriptstyle\div}$\!\!\!\!&${\scriptstyle 0.325}$&${\scriptstyle 181}$ &  ${\scriptstyle \pm 12}$ &  ${\scriptstyle +19/\!-\!24}$ &  ${\scriptstyle +42/\!-\!36}$ & ${\scriptstyle 0.71}$\\
${\scriptstyle 0.325}$&\!\!\!\!${\scriptstyle\div}$\!\!\!\!&${\scriptstyle 0.55}$&${\scriptstyle 287}$ &  ${\scriptstyle \pm 12}$ &  ${\scriptstyle +29/\!-\!42}$ &  ${\scriptstyle +52/\!-\!47}$ & ${\scriptstyle 0.78}$\\
${\scriptstyle 0.55}$&\!\!\!\!${\scriptstyle\div}$\!\!\!\!&${\scriptstyle 0.775}$&${\scriptstyle 423}$ &  ${\scriptstyle \pm 15}$ &  ${\scriptstyle +45/\!-\!49}$ &  ${\scriptstyle +57/\!-\!58}$ & ${\scriptstyle 1.2}$\\
${\scriptstyle 0.775}$&\!\!\!\!${\scriptstyle\div}$\!\!\!\!&${\scriptstyle 1}$&${\scriptstyle 428}$ &  ${\scriptstyle \pm 14}$ &  ${\scriptstyle +44/\!-\!50}$ &  ${\scriptstyle +45/\!-\!48}$ & ${\scriptstyle 0.71}$\\
\hline
\hline
&&&&&&& \\[-10pt]
\multicolumn{3}{|c|}{$E_{T}^{\rm jet1}\,\,[\!~\mbox{GeV}]$} & $\mbox{d}\sigma/\mbox{d}E_{T}^{\rm jet1}\,\,[\!~\mbox{pb}~\mbox{GeV}^{-1}]$ & $\delta_{\mbox{\tiny stat}}$ & $\delta_{\mbox{\tiny uncorr}}$ & $\delta_{\mbox{\tiny corr}}$ & $1+\delta_{\mbox{\tiny hadr}}$ \\[0.1cm]
\hline
${\scriptstyle 5}$&\!\!\!\!${\scriptstyle\div}$\!\!\!\!&${\scriptstyle 7}$&${\scriptstyle 104}$ &  ${\scriptstyle \pm 3}$ &  ${\scriptstyle +9.7/\!-\!12}$ &  ${\scriptstyle +13/\!-\!14}$ & ${\scriptstyle 0.82}$\\
${\scriptstyle 7}$&\!\!\!\!${\scriptstyle\div}$\!\!\!\!&${\scriptstyle 10}$&${\scriptstyle 25.3}$ &  ${\scriptstyle \pm 0.9}$ &  ${\scriptstyle +2.5/\!-\!3.2}$ &  ${\scriptstyle +3.7/\!-\!3.5}$ & ${\scriptstyle 0.94}$\\
${\scriptstyle 10}$&\!\!\!\!${\scriptstyle\div}$\!\!\!\!&${\scriptstyle 14}$&${\scriptstyle 1.96}$ &  ${\scriptstyle \pm 0.16}$ &  ${\scriptstyle +0.29/\!-\!0.25}$ &  ${\scriptstyle +0.36/\!-\!0.36}$ & ${\scriptstyle 0.90}$\\
\hline
\hline
&&&&&&& \\[-10pt]
\multicolumn{3}{|c|}{$\mbox{log}(x_{\pom})$} & $\mbox{d}\sigma/\mbox{d}\mbox{log}(x_{\pom})\,\,[\mbox{pb}]$ & $\delta_{\mbox{\tiny stat}}$ & $\delta_{\mbox{\tiny uncorr}}$ & $\delta_{\mbox{\tiny corr}}$ & $1+\delta_{\mbox{\tiny hadr}}$ \\[0.1cm]
\hline
${\scriptstyle -2.3}$&\!\!\!\!${\scriptstyle\div}$\!\!\!\!&${\scriptstyle -2.1}$&${\scriptstyle 92}$ &  ${\scriptstyle \pm 7}$ &  ${\scriptstyle +8/\!-\!11}$ &  ${\scriptstyle +13/\!-\!14}$ & ${\scriptstyle 0.93}$\\
${\scriptstyle -2.1}$&\!\!\!\!${\scriptstyle\div}$\!\!\!\!&${\scriptstyle -1.9}$&${\scriptstyle 238}$ &  ${\scriptstyle \pm 13}$ &  ${\scriptstyle +22/\!-\!30}$ &  ${\scriptstyle +31/\!-\!27}$ & ${\scriptstyle 0.90}$\\
${\scriptstyle -1.9}$&\!\!\!\!${\scriptstyle\div}$\!\!\!\!&${\scriptstyle -1.7}$&${\scriptstyle 466}$ &  ${\scriptstyle \pm 16}$ &  ${\scriptstyle +50/\!-\!60}$ &  ${\scriptstyle +60/\!-\!61}$ & ${\scriptstyle 0.84}$\\
${\scriptstyle -1.7}$&\!\!\!\!${\scriptstyle\div}$\!\!\!\!&${\scriptstyle -1.523}$&${\scriptstyle 734}$ &  ${\scriptstyle \pm 23}$ &  ${\scriptstyle +78/\!-\!85}$ &  ${\scriptstyle +11/\!-\!10}$ & ${\scriptstyle 0.82}$\\
\hline
\hline
&&&&&&& \\[-10pt]
\multicolumn{3}{|c|}{$z^{\rm jets}_{\pom}$} & $\mbox{d}\sigma/\mbox{d}z^{\rm jets}_{\pom}\,\,[\mbox{pb}]$ & $\delta_{\mbox{\tiny stat}}$ & $\delta_{\mbox{\tiny uncorr}}$ & $\delta_{\mbox{\tiny corr}}$ & $1+\delta_{\mbox{\tiny hadr}}$ \\[0.1cm]
\hline
${\scriptstyle 0.14}$&\!\!\!\!${\scriptstyle\div}$\!\!\!\!&${\scriptstyle 0.36}$&${\scriptstyle 471}$ &  ${\scriptstyle \pm 19}$ &  ${\scriptstyle +54/\!-\!66}$ &  ${\scriptstyle +68/\!-\!67}$ & ${\scriptstyle 0.79}$\\
${\scriptstyle 0.36}$&\!\!\!\!${\scriptstyle\div}$\!\!\!\!&${\scriptstyle 0.58}$&${\scriptstyle 460}$ &  ${\scriptstyle \pm 15}$ &  ${\scriptstyle +54/\!-\!63}$ &  ${\scriptstyle +60/\!-\!59}$ & ${\scriptstyle 0.86}$\\
${\scriptstyle 0.58}$&\!\!\!\!${\scriptstyle\div}$\!\!\!\!&${\scriptstyle 0.8}$&${\scriptstyle 374}$ &  ${\scriptstyle \pm 12}$ &  ${\scriptstyle +42/\!-\!47}$ &  ${\scriptstyle +49/\!-\!49}$ & ${\scriptstyle 0.92}$\\
${\scriptstyle 0.8}$&\!\!\!\!${\scriptstyle\div}$\!\!\!\!&${\scriptstyle 1}$&${\scriptstyle 151}$ &  ${\scriptstyle \pm 12}$ & ${\scriptstyle +19/\!-\!18}$ &  ${\scriptstyle +21/\!-\!22}$ & $ - $\\
\hline
\hline
&&&&&&& \\[-10pt]
\multicolumn{3}{|c|}{$\left<\eta^{\rm jets}\right>$} & $\mbox{d}\sigma/\mbox{d}\left<\eta^{\rm jets}\right>\,\,[\mbox{pb}]$ & $\delta_{\mbox{\tiny stat}}$ & $\delta_{\mbox{\tiny uncorr}}$ & $\delta_{\mbox{\tiny corr}}$ & $1+\delta_{\mbox{\tiny hadr}}$ \\[0.1cm]
\hline
${\scriptstyle -0.7}$&\!\!\!\!${\scriptstyle\div}$\!\!\!\!&${\scriptstyle -0.4}$&${\scriptstyle 181}$ &  ${\scriptstyle \pm 1}$ &  ${\scriptstyle +15/\!-\!20}$ &  ${\scriptstyle +29/\!-\!29}$ & ${\scriptstyle 0.79}$\\
${\scriptstyle -0.4}$&\!\!\!\!${\scriptstyle\div}$\!\!\!\!&${\scriptstyle -0.1}$&${\scriptstyle 280}$ &  ${\scriptstyle \pm 10}$ &  ${\scriptstyle +26/\!-\!31}$ &  ${\scriptstyle +36/\!-\!37}$ & ${\scriptstyle 0.93}$\\
${\scriptstyle -0.1}$&\!\!\!\!${\scriptstyle\div}$\!\!\!\!&${\scriptstyle 0.2}$&${\scriptstyle 224}$ &  ${\scriptstyle \pm 10}$ &  ${\scriptstyle +24/\!-\!30}$ &  ${\scriptstyle +27/\!-\!28}$ & ${\scriptstyle 0.93}$\\
${\scriptstyle 0.2}$&\!\!\!\!${\scriptstyle\div}$\!\!\!\!&${\scriptstyle 0.5}$&${\scriptstyle 153}$ &  ${\scriptstyle \pm 7}$ &  ${\scriptstyle +17/\!-\!18}$ &  ${\scriptstyle +20/\!-\!19}$ & ${\scriptstyle 0.80}$\\
${\scriptstyle 0.5}$&\!\!\!\!${\scriptstyle\div}$\!\!\!\!&${\scriptstyle 0.8}$&${\scriptstyle 75}$ &  ${\scriptstyle \pm 5}$ &  ${\scriptstyle +8/\!-\!10}$ &  ${\scriptstyle +10/\!-\!10}$ & ${\scriptstyle 0.75}$\\
\hline
\hline
&&&&&&& \\[-10pt]
\multicolumn{3}{|c|}{$\left|\Delta \eta^{\rm jets}\right|$} & $\mbox{d}\sigma/\mbox{d}\left|\Delta \eta^{\rm jets}\right|\,\,[\mbox{pb}]$ & $\delta_{\mbox{\tiny stat}}$ & $\delta_{\mbox{\tiny uncorr}}$ & $\delta_{\mbox{\tiny corr}}$ & $1+\delta_{\mbox{\tiny hadr}}$ \\[0.1cm]
\hline
${\scriptstyle 0}$&\!\!\!\!${\scriptstyle\div}$\!\!\!\!&${\scriptstyle 0.4}$&${\scriptstyle 216}$ &  ${\scriptstyle \pm 8}$ &  ${\scriptstyle +21/\!-\!22}$ &  ${\scriptstyle +32/\!-\!31}$ & ${\scriptstyle 0.81}$\\
${\scriptstyle 0.4}$&\!\!\!\!${\scriptstyle\div}$\!\!\!\!&${\scriptstyle 0.8}$&${\scriptstyle 195}$ &  ${\scriptstyle \pm 8}$ &  ${\scriptstyle +18/\!-\!22}$ &  ${\scriptstyle +26/\!-\!27}$ & ${\scriptstyle 0.83}$\\
${\scriptstyle 0.8}$&\!\!\!\!${\scriptstyle\div}$\!\!\!\!&${\scriptstyle 1.2}$&${\scriptstyle 149}$ &  ${\scriptstyle \pm 6}$ &  ${\scriptstyle +12/\!-\!18}$ &  ${\scriptstyle +20/\!-\!19}$ & ${\scriptstyle 0.85}$\\
${\scriptstyle 1.2}$&\!\!\!\!${\scriptstyle\div}$\!\!\!\!&${\scriptstyle 1.6}$&${\scriptstyle 96}$ &  ${\scriptstyle \pm 6}$ &  ${\scriptstyle +11/\!-\!14}$ &  ${\scriptstyle +13/\!-\!12}$ & ${\scriptstyle 0.93}$\\
${\scriptstyle 1.6}$&\!\!\!\!${\scriptstyle\div}$\!\!\!\!&${\scriptstyle 2}$&${\scriptstyle 51.7}$ &  ${\scriptstyle \pm 3.5}$ &  ${\scriptstyle +5.5/\!-\!6.2}$ &  ${\scriptstyle +6.8/\!-\!6.9}$ & ${\scriptstyle 0.80}$\\
\hline
\hline
&&&&&&& \\[-10pt]
\multicolumn{3}{|c|}{$W\,\,[\!~\mbox{GeV}]$} & $\mbox{d}\sigma/\mbox{d}W\,\,[\!~\mbox{pb}~\mbox{GeV}^{-1}]$ & $\delta_{\mbox{\tiny stat}}$ & $\delta_{\mbox{\tiny uncorr}}$ & $\delta_{\mbox{\tiny corr}}$ & $1+\delta_{\mbox{\tiny hadr}}$ \\[0.1cm]
\hline
${\scriptstyle 175}$&\!\!\!\!${\scriptstyle\div}$\!\!\!\!&${\scriptstyle 191}$&${\scriptstyle 4.50}$ &  ${\scriptstyle \pm 0.21}$ &  ${\scriptstyle +0.51/\!-\!0.70}$ &  ${\scriptstyle +0.63/\!-\!0.71}$ & ${\scriptstyle 0.80}$\\
${\scriptstyle 191}$&\!\!\!\!${\scriptstyle\div}$\!\!\!\!&${\scriptstyle 207}$&${\scriptstyle 4.14}$ &  ${\scriptstyle \pm 0.17}$ &  ${\scriptstyle +0.70/\!-\!0.40}$ &  ${\scriptstyle +0.58/\!-\!0.62}$ & ${\scriptstyle 0.84}$\\
${\scriptstyle 207}$&\!\!\!\!${\scriptstyle\div}$\!\!\!\!&${\scriptstyle 223}$&${\scriptstyle 3.67}$ &  ${\scriptstyle \pm 0.14}$ &  ${\scriptstyle +0.31/\!-\!0.53}$ &  ${\scriptstyle +0.52/\!-\!0.53}$ & ${\scriptstyle 0.88}$\\
${\scriptstyle 223}$&\!\!\!\!${\scriptstyle\div}$\!\!\!\!&${\scriptstyle 239}$&${\scriptstyle 3.30}$ &  ${\scriptstyle \pm 0.16}$ &  ${\scriptstyle +0.31/\!-\!0.37}$ &  ${\scriptstyle +0.43/\!-\!0.38}$ & ${\scriptstyle 0.86}$\\
${\scriptstyle 239}$&\!\!\!\!${\scriptstyle\div}$\!\!\!\!&${\scriptstyle 255}$&${\scriptstyle 2.46}$ &  ${\scriptstyle \pm 0.14}$ &  ${\scriptstyle +0.25/\!-\!0.29}$ &  ${\scriptstyle +0.31/\!-\!0.28}$ & ${\scriptstyle 0.85}$\\
\hline
\hline
&&&&&&& \\[-10pt]
\multicolumn{3}{|c|}{$M_{12}\,\,[\!~\mbox{GeV}]$} & $\mbox{d}\sigma/\mbox{d}M_{12}\,\,[\!~\mbox{pb}~\mbox{GeV}^{-1}]$ & $\delta_{\mbox{\tiny stat}}$ & $\delta_{\mbox{\tiny uncorr}}$ & $\delta_{\mbox{\tiny corr}}$ & $1+\delta_{\mbox{\tiny hadr}}$ \\[0.1cm]
\hline
${\scriptstyle 9}$&\!\!\!\!${\scriptstyle\div}$\!\!\!\!&${\scriptstyle 14}$&${\scriptstyle 38.2}$ &  ${\scriptstyle \pm 1.0}$ &  ${\scriptstyle +3.3/\!-\!4.3}$ &  ${\scriptstyle +5.1/\!-\!5.0}$ & ${\scriptstyle 0.82}$\\
${\scriptstyle 14}$&\!\!\!\!${\scriptstyle\div}$\!\!\!\!&${\scriptstyle 19}$&${\scriptstyle 15.4}$ &  ${\scriptstyle \pm 0.6}$ &  ${\scriptstyle +1.7/\!-\!2.0}$ &  ${\scriptstyle +1.9/\!-\!2.0}$ & ${\scriptstyle 0.88}$\\
${\scriptstyle 19}$&\!\!\!\!${\scriptstyle\div}$\!\!\!\!&${\scriptstyle 29}$&${\scriptstyle 2.12}$ &  ${\scriptstyle \pm 0.14}$ &  ${\scriptstyle +0.26/\!-\!0.23}$ &  ${\scriptstyle +0.43/\!-\!0.35}$ & ${\scriptstyle 0.97}$\\
\hline
\hline
&&&&&&& \\[-10pt]
\multicolumn{3}{|c|}{$M_{X}\,\,[\!~\mbox{GeV}]$} & $\mbox{d}\sigma/\mbox{d}M_{X}\,\,[\!~\mbox{pb}~\mbox{GeV}^{-1}]$ & $\delta_{\mbox{\tiny stat}}$ & $\delta_{\mbox{\tiny uncorr}}$ & $\delta_{\mbox{\tiny corr}}$ & $1+\delta_{\mbox{\tiny hadr}}$ \\[0.1cm]
\hline
${\scriptstyle 10}$&\!\!\!\!${\scriptstyle\div}$\!\!\!\!&${\scriptstyle 17.5}$&${\scriptstyle 2.61}$ &  ${\scriptstyle \pm 0.20}$ &  ${\scriptstyle +0.28/\!-\!0.30}$ &  ${\scriptstyle \pm 0.45}$ & ${\scriptstyle 0.94}$\\
${\scriptstyle 17.5}$&\!\!\!\!${\scriptstyle\div}$\!\!\!\!&${\scriptstyle 25}$&${\scriptstyle 10.1}$ &  ${\scriptstyle \pm 0.4}$ &  ${\scriptstyle +1.0/\!-\!1.3}$ &  ${\scriptstyle \pm 1.3}$ & ${\scriptstyle 0.87}$\\
${\scriptstyle 25}$&\!\!\!\!${\scriptstyle\div}$\!\!\!\!&${\scriptstyle 40}$&${\scriptstyle 12.8}$ &  ${\scriptstyle \pm 0.3}$ &  ${\scriptstyle +1.2/\!-\!1.5}$ &  ${\scriptstyle \pm 1.7}$ & ${\scriptstyle 0.83}$\\
\hline
\end{tabular}
\caption[ab]{Bin averaged hadron level differential cross sections for diffractive dijet photoproduction. For each data point, the absolute statistical, uncorrelated and correlated systematic uncertainties and the hadronisation correction factors applied to the NLO calculations are given as $\delta_{\mbox{\tiny stat}}$, $\delta_{\mbox{\tiny uncorr}}$ and $\delta_{\mbox{\tiny corr}}$ and $1+\delta_{\mbox{\tiny hadr}}$, respectively.}
\label{table1D}
}
\end{center}
\end{table}

\begin{table}[htb!]
\begin{center}
{\footnotesize
\begin{tabular}{|ccc|c|c|c|c|c|c|}
\hline
&&&&&&&& \\[-10pt]
\multicolumn{3}{|c|}{$x^{\rm jets}_{\gamma}$} & $\mbox{data}/\mbox{theory}$ & $\delta_{\mbox{\tiny stat}}$ & $\delta_{\mbox{\tiny uncorr}}$ & $\delta_{\mbox{\tiny corr}}$ & $\delta_{\mbox{\tiny DPDF}}$ & $\delta_{\mbox{\tiny scal}}$ \\[0.1cm]
\hline
${\scriptstyle 0.1}$&\!\!\!\!${\scriptstyle \div}$\!\!\!\!&${\scriptstyle 0.325}$&${\scriptstyle 0.62}$ &  ${\scriptstyle \pm 0.04}$ &  ${\scriptstyle+0.06/-0.08}$ &  ${\scriptstyle+0.14/-0.12}$ &  ${\scriptstyle+0.12/-0.06}$ &  ${\scriptstyle+0.28/-0.23}$ \\
${\scriptstyle 0.325}$&\!\!\!\!${\scriptstyle \div}$\!\!\!\!&${\scriptstyle 0.55}$&${\scriptstyle 0.55}$ &  ${\scriptstyle \pm 0.02}$ &  ${\scriptstyle+0.06/-0.08}$ &  ${\scriptstyle+0.01/-0.09}$ &  ${\scriptstyle+0.09/-0.05}$ &  ${\scriptstyle+0.16/-0.15}$ \\
${\scriptstyle 0.55}$&\!\!\!\!${\scriptstyle \div}$\!\!\!\!&${\scriptstyle 0.775}$&${\scriptstyle 0.57}$ &  ${\scriptstyle \pm 0.02}$ &  ${\scriptstyle+0.06/-0.07}$ &  ${\scriptstyle+0.08/-0.08}$ &  ${\scriptstyle+0.09/-0.05}$ &  ${\scriptstyle+0.11/-0.10}$ \\
${\scriptstyle 0.775}$&\!\!\!\!${\scriptstyle \div}$\!\!\!\!&${\scriptstyle 1}$&${\scriptstyle 0.62}$ &  ${\scriptstyle \pm 0.02}$ &  ${\scriptstyle+0.06/-0.07}$ &  ${\scriptstyle+0.07/-0.07}$ &  ${\scriptstyle+0.10/-0.06}$ &  ${\scriptstyle+0.10/-0.01}$ \\
\hline
\hline
&&&&&&&& \\[-10pt]
\multicolumn{3}{|c|}{$E_{T}^{\rm jet1}\,\,[\!~\mbox{GeV}]$} & $\mbox{data}/\mbox{theory}$ & $\delta_{\mbox{\tiny stat}}$ & $\delta_{\mbox{\tiny uncorr}}$ & $\delta_{\mbox{\tiny corr}}$ & $\delta_{\mbox{\tiny DPDF}}$ & $\delta_{\mbox{\tiny scal}}$ \\[0.1cm]
\hline
${\scriptstyle 5}$&\!\!\!\!${\scriptstyle \div}$\!\!\!\!&${\scriptstyle 7}$&${\scriptstyle 0.54}$ &  ${\scriptstyle \pm 0.01}$ &  ${\scriptstyle+0.05/-0.06}$ &  ${\scriptstyle+0.07/-0.07}$ &  ${\scriptstyle+0.09/-0.05}$ &  ${\scriptstyle+0.13/-0.12}$ \\
${\scriptstyle 7}$&\!\!\!\!${\scriptstyle \div}$\!\!\!\!&${\scriptstyle 10}$&${\scriptstyle 0.66}$ &  ${\scriptstyle \pm 0.02}$ &  ${\scriptstyle+0.07/-0.08}$ &  ${\scriptstyle+0.10/-0.09}$ &  ${\scriptstyle+0.11/-0.07}$ &  ${\scriptstyle+0.18/-0.17}$ \\
${\scriptstyle 10}$&\!\!\!\!${\scriptstyle \div}$\!\!\!\!&${\scriptstyle 14}$&${\scriptstyle 0.71}$ &  ${\scriptstyle \pm 0.06}$ & ${\scriptstyle+0.10/-0.09}$ &  ${\scriptstyle+0.13/-0.13}$ &  ${\scriptstyle+0.13/-0.08}$ &  ${\scriptstyle+0.20/-0.18}$ \\
\hline
\hline
&&&&&&&& \\[-10pt]
\multicolumn{3}{|c|}{$\mbox{log}(x_{\pom})$} & $\mbox{data}/\mbox{theory}$ & $\delta_{\mbox{\tiny stat}}$ & $\delta_{\mbox{\tiny uncorr}}$ & $\delta_{\mbox{\tiny corr}}$ & $\delta_{\mbox{\tiny DPDF}}$ & $\delta_{\mbox{\tiny scal}}$ \\[0.1cm]
\hline
${\scriptstyle -2.3}$&\!\!\!\!${\scriptstyle \div}$\!\!\!\!&${\scriptstyle -2.1}$&${\scriptstyle 0.50}$ &  ${\scriptstyle \pm 0.08}$ &  ${\scriptstyle+0.04/-0.06}$ &  ${\scriptstyle+0.07/-0.08}$ &  ${\scriptstyle+0.07/-0.05}$ &  ${\scriptstyle+0.16/-0.14}$ \\
${\scriptstyle -2.1}$&\!\!\!\!${\scriptstyle \div}$\!\!\!\!&${\scriptstyle -1.9}$&${\scriptstyle 0.54}$ &  ${\scriptstyle \pm 0.03}$ &  ${\scriptstyle+0.05/-0.07}$ &  ${\scriptstyle+0.07/-0.06}$ &  ${\scriptstyle+0.08/-0.05}$ &  ${\scriptstyle+0.15/-0.14}$ \\
${\scriptstyle -1.9}$&\!\!\!\!${\scriptstyle \div}$\!\!\!\!&${\scriptstyle -1.7}$&${\scriptstyle 0.60}$ &  ${\scriptstyle \pm 0.02}$ &  ${\scriptstyle+0.06/-0.08}$ &  ${\scriptstyle+0.08/-0.08}$ &  ${\scriptstyle+0.10/-0.06}$ &  ${\scriptstyle +0.14/-0.14}$ \\
${\scriptstyle -1.7}$&\!\!\!\!${\scriptstyle \div}$\!\!\!\!&${\scriptstyle -1.523}$&${\scriptstyle 0.59}$ &  ${\scriptstyle \pm 0.02}$ &  ${\scriptstyle+0.06/-0.07}$ &  ${\scriptstyle+0.09/-0.08}$ &  ${\scriptstyle+0.11/-0.06}$ &  ${\scriptstyle +0.13/-0.13}$ \\
\hline
\hline
&&&&&&&& \\[-10pt]
\multicolumn{3}{|c|}{$z^{\rm jets}_{\pom}$} & $\mbox{data}/\mbox{theory}$ & $\delta_{\mbox{\tiny stat}}$ & $\delta_{\mbox{\tiny uncorr}}$ & $\delta_{\mbox{\tiny corr}}$ & $\delta_{\mbox{\tiny DPDF}}$ & $\delta_{\mbox{\tiny scal}}$ \\[0.1cm]
\hline
${\scriptstyle 0.14}$&\!\!\!\!${\scriptstyle \div}$\!\!\!\!&${\scriptstyle 0.36}$&${\scriptstyle 0.62}$ &  ${\scriptstyle \pm 0.03}$ &  ${\scriptstyle+0.07/-0.09}$ &  ${\scriptstyle+0.09/-0.09}$ &  ${\scriptstyle+0.10/-0.06}$ &  ${\scriptstyle+0.11/-0.12}$ \\
${\scriptstyle 0.36}$&\!\!\!\!${\scriptstyle \div}$\!\!\!\!&${\scriptstyle 0.58}$&${\scriptstyle 0.55}$ &  ${\scriptstyle \pm 0.02}$ &  ${\scriptstyle+0.07/-0.08}$ &  ${\scriptstyle+0.07/-0.07}$ &  ${\scriptstyle+0.09/-0.05}$ &  ${\scriptstyle+0.14/-0.13}$ \\
${\scriptstyle 0.58}$&\!\!\!\!${\scriptstyle \div}$\!\!\!\!&${\scriptstyle 0.8}$&${\scriptstyle 0.57}$ &  ${\scriptstyle \pm 0.02}$ &  ${\scriptstyle+0.06/-0.07}$ &  ${\scriptstyle+0.07/-0.07}$ &  ${\scriptstyle+0.10/-0.06}$ &  ${\scriptstyle+0.20/-0.16}$ \\
\hline
\hline
&&&&&&&& \\[-10pt]
\multicolumn{3}{|c|}{$\left<\eta^{\rm jets}\right>$} & $\mbox{data}/\mbox{theory}$ & $\delta_{\mbox{\tiny stat}}$ & $\delta_{\mbox{\tiny uncorr}}$ & $\delta_{\mbox{\tiny corr}}$ & $\delta_{\mbox{\tiny DPDF}}$ & $\delta_{\mbox{\tiny scal}}$ \\[0.1cm]
\hline
${\scriptstyle -0.7}$&\!\!\!\!${\scriptstyle \div}$\!\!\!\!&${\scriptstyle -0.4}$&${\scriptstyle 0.58}$ &  ${\scriptstyle \pm 0.03}$ &  ${\scriptstyle+0.05/-0.06}$ &  ${\scriptstyle+0.09/-0.09}$ &  ${\scriptstyle+0.09/-0.05}$ &  ${\scriptstyle +0.10/-0.10}$ \\
${\scriptstyle -0.4}$&\!\!\!\!${\scriptstyle \div}$\!\!\!\!&${\scriptstyle -0.1}$&${\scriptstyle 0.62}$ &  ${\scriptstyle \pm 0.02}$ &  ${\scriptstyle+0.06/-0.07}$ &  ${\scriptstyle+0.08/-0.08}$ &  ${\scriptstyle+0.10/-0.06}$ &  ${\scriptstyle +0.12/-0.12}$ \\
${\scriptstyle -0.1}$&\!\!\!\!${\scriptstyle \div}$\!\!\!\!&${\scriptstyle 0.2}$&${\scriptstyle 0.56}$ &  ${\scriptstyle \pm 0.02}$ &  ${\scriptstyle+0.06/-0.07}$ &  ${\scriptstyle+0.07/-0.07}$ &  ${\scriptstyle+0.10/-0.05}$ &  ${\scriptstyle+0.14/-0.12}$ \\
${\scriptstyle 0.2}$&\!\!\!\!${\scriptstyle \div}$\!\!\!\!&${\scriptstyle 0.5}$&${\scriptstyle 0.62}$ &  ${\scriptstyle \pm 0.03}$ &  ${\scriptstyle+0.07/-0.07}$ &  ${\scriptstyle+0.08/-0.08}$ &  ${\scriptstyle+0.11/-0.06}$ &  ${\scriptstyle+0.19/-0.17}$ \\
${\scriptstyle 0.5}$&\!\!\!\!${\scriptstyle \div}$\!\!\!\!&${\scriptstyle 0.8}$&${\scriptstyle 0.57}$ &  ${\scriptstyle \pm 0.04}$ &  ${\scriptstyle+0.06/-0.08}$ &  ${\scriptstyle+0.08/-0.07}$ &  ${\scriptstyle+0.11/-0.06}$ &  ${\scriptstyle+0.23/-0.18}$ \\
\hline
\hline
&&&&&&&& \\[-10pt]
\multicolumn{3}{|c|}{$\left|\Delta \eta^{\rm jets}\right|$} & $\mbox{data}/\mbox{theory}$ & $\delta_{\mbox{\tiny stat}}$ & $\delta_{\mbox{\tiny uncorr}}$ & $\delta_{\mbox{\tiny corr}}$ & $\delta_{\mbox{\tiny DPDF}}$ & $\delta_{\mbox{\tiny scal}}$ \\[0.1cm]
\hline
${\scriptstyle 0}$&\!\!\!\!${\scriptstyle \div}$\!\!\!\!&${\scriptstyle 0.4}$&${\scriptstyle 0.59}$ &  ${\scriptstyle \pm 0.02}$ &  ${\scriptstyle+0.06/-0.06}$ &  ${\scriptstyle+0.09/-0.09}$ &  ${\scriptstyle+0.09/-0.06}$ &  ${\scriptstyle +0.12/-0.12}$ \\
${\scriptstyle 0.4}$&\!\!\!\!${\scriptstyle \div}$\!\!\!\!&${\scriptstyle 0.8}$&${\scriptstyle 0.62}$ &  ${\scriptstyle \pm 0.03}$ &  ${\scriptstyle+0.06/-0.07}$ &  ${\scriptstyle+0.08/-0.09}$ &  ${\scriptstyle+0.10/-0.06}$ &  ${\scriptstyle+0.14/-0.13}$ \\
${\scriptstyle 0.8}$&\!\!\!\!${\scriptstyle \div}$\!\!\!\!&${\scriptstyle 1.2}$&${\scriptstyle 0.61}$ &  ${\scriptstyle \pm 0.03}$ &  ${\scriptstyle+0.05/-0.07}$ &  ${\scriptstyle+0.08/-0.08}$ &  ${\scriptstyle+0.11/-0.06}$ &  ${\scriptstyle+0.15/-0.14}$ \\
${\scriptstyle 1.2}$&\!\!\!\!${\scriptstyle \div}$\!\!\!\!&${\scriptstyle 1.6}$&${\scriptstyle 0.51}$ &  ${\scriptstyle \pm 0.03}$ &  ${\scriptstyle+0.06/-0.07}$ &  ${\scriptstyle+0.07/-0.07}$ &  ${\scriptstyle+0.09/-0.05}$ &  ${\scriptstyle+0.14/-0.13}$ \\
${\scriptstyle 1.6}$&\!\!\!\!${\scriptstyle \div}$\!\!\!\!&${\scriptstyle 2}$&${\scriptstyle 0.55}$ &  ${\scriptstyle \pm 0.04}$ &  ${\scriptstyle+0.06/-0.07}$ &  ${\scriptstyle+0.07/-0.07}$ &  ${\scriptstyle+0.10/-0.05}$ &  ${\scriptstyle+0.17/-0.16}$ \\
\hline
\hline
&&&&&&&& \\[-10pt]
\multicolumn{3}{|c|}{$W\,\,[\!~\mbox{GeV}]$} & $\mbox{data}/\mbox{theory}$ & $\delta_{\mbox{\tiny stat}}$ & $\delta_{\mbox{\tiny uncorr}}$ & $\delta_{\mbox{\tiny corr}}$ & $\delta_{\mbox{\tiny DPDF}}$ & $\delta_{\mbox{\tiny scal}}$ \\[0.1cm]
\hline
${\scriptstyle 175}$&\!\!\!\!${\scriptstyle \div}$\!\!\!\!&${\scriptstyle 191}$&${\scriptstyle 0.59}$ &  ${\scriptstyle \pm 0.03}$ &  ${\scriptstyle+0.07/-0.09}$ &  ${\scriptstyle+0.08/-0.09}$ &  ${\scriptstyle+0.10/-0.06}$ &  ${\scriptstyle+0.15/-0.14}$ \\
${\scriptstyle 191}$&\!\!\!\!${\scriptstyle \div}$\!\!\!\!&${\scriptstyle 207}$&${\scriptstyle 0.57}$ &  ${\scriptstyle \pm 0.02}$ &  ${\scriptstyle+0.10/-0.06}$ &  ${\scriptstyle+0.08/-0.09}$ &  ${\scriptstyle+0.10/-0.05}$ &  ${\scriptstyle+0.14/-0.13}$ \\
${\scriptstyle 207}$&\!\!\!\!${\scriptstyle \div}$\!\!\!\!&${\scriptstyle 223}$&${\scriptstyle 0.57}$ &  ${\scriptstyle \pm 0.02}$ &  ${\scriptstyle+0.05/-0.08}$ &  ${\scriptstyle+0.08/-0.08}$ &  ${\scriptstyle+0.10/-0.05}$ &  ${\scriptstyle +0.13/-0.13}$ \\
${\scriptstyle 223}$&\!\!\!\!${\scriptstyle \div}$\!\!\!\!&${\scriptstyle 239}$&${\scriptstyle 0.62}$ &  ${\scriptstyle \pm 0.03}$ &  ${\scriptstyle+0.06/-0.07}$ &  ${\scriptstyle+0.08/-0.07}$ &  ${\scriptstyle+0.10/-0.06}$ &  ${\scriptstyle+0.15/-0.14}$ \\
${\scriptstyle 239}$&\!\!\!\!${\scriptstyle \div}$\!\!\!\!&${\scriptstyle 255}$&${\scriptstyle 0.56}$ &  ${\scriptstyle \pm 0.03}$ &  ${\scriptstyle+0.06/-0.07}$ &  ${\scriptstyle+0.07/-0.06}$ &  ${\scriptstyle+0.09/-0.06}$ &  ${\scriptstyle+0.14/-0.13}$ \\
\hline
\hline
&&&&&&&& \\[-10pt]
\multicolumn{3}{|c|}{$M_{12}\,\,[\!~\mbox{GeV}]$} & $\mbox{data}/\mbox{theory}$ & $\delta_{\mbox{\tiny stat}}$ & $\delta_{\mbox{\tiny uncorr}}$ & $\delta_{\mbox{\tiny corr}}$ & $\delta_{\mbox{\tiny DPDF}}$ & $\delta_{\mbox{\tiny scal}}$ \\[0.1cm]
\hline
${\scriptstyle 9}$&\!\!\!\!${\scriptstyle \div}$\!\!\!\!&${\scriptstyle 14}$&${\scriptstyle 0.55}$ &  ${\scriptstyle \pm 0.01}$ &  ${\scriptstyle+0.05/-0.06}$ &  ${\scriptstyle+0.07/-0.07}$ &  ${\scriptstyle+0.09/-0.05}$ &  ${\scriptstyle +0.15/-0.15}$ \\
${\scriptstyle 14}$&\!\!\!\!${\scriptstyle \div}$\!\!\!\!&${\scriptstyle 19}$&${\scriptstyle 0.63}$ &  ${\scriptstyle \pm 0.03}$ &  ${\scriptstyle+0.07/-0.08}$ &  ${\scriptstyle+0.08/-0.08}$ &  ${\scriptstyle+0.11/-0.06}$ &  ${\scriptstyle+0.10/-0.08}$ \\
${\scriptstyle 19}$&\!\!\!\!${\scriptstyle \div}$\!\!\!\!&${\scriptstyle 29}$&${\scriptstyle 0.62}$ &  ${\scriptstyle \pm 0.04}$ &  ${\scriptstyle+0.08/-0.07}$ &  ${\scriptstyle+0.13/-0.10}$ &  ${\scriptstyle+0.11/-0.07}$ &  ${\scriptstyle+0.10/-0.05}$ \\
\hline
\hline
&&&&&&&& \\[-10pt]
\multicolumn{3}{|c|}{$M_{X}\,\,[\!~\mbox{GeV}]$} & $\mbox{data}/\mbox{theory}$ & $\delta_{\mbox{\tiny stat}}$ & $\delta_{\mbox{\tiny uncorr}}$ & $\delta_{\mbox{\tiny corr}}$ & $\delta_{\mbox{\tiny DPDF}}$ & $\delta_{\mbox{\tiny scal}}$ \\[0.1cm]
\hline
${\scriptstyle 10}$&\!\!\!\!${\scriptstyle \div}$\!\!\!\!&${\scriptstyle 17.5}$&${\scriptstyle 0.51}$ &  ${\scriptstyle \pm 0.04}$ &  ${\scriptstyle+0.06/-0.06}$ &  ${\scriptstyle+0.09/-0.09}$ &  ${\scriptstyle+0.08/-0.05}$ &  ${\scriptstyle+0.17/-0.15}$ \\
${\scriptstyle 17.5}$&\!\!\!\!${\scriptstyle \div}$\!\!\!\!&${\scriptstyle 25}$&${\scriptstyle 0.55}$ &  ${\scriptstyle \pm 0.02}$ &  ${\scriptstyle+0.05/-0.07}$ &  ${\scriptstyle+0.07/-0.07}$ &  ${\scriptstyle+0.09/-0.05}$ &  ${\scriptstyle+0.14/-0.13}$ \\
${\scriptstyle 25}$&\!\!\!\!${\scriptstyle \div}$\!\!\!\!&${\scriptstyle 40}$&${\scriptstyle 0.60}$ &  ${\scriptstyle \pm 0.02}$ &  ${\scriptstyle+0.68/-0.07}$ &  ${\scriptstyle+0.08/-0.08}$ &  ${\scriptstyle+0.11/-0.06}$ &  ${\scriptstyle+0.14/-0.13}$ \\
\hline
\end{tabular}
\caption{Ratios of differential cross sections for diffractive dijet photoproduction to NLO QCD calculations obtained in the FR framework with the H1 2006 Fit B DPDFs, corrected for hadronisation. For each ratio point, the absolute statistical, uncorrelated and correlated systematic uncertainties from the data and the absolute DPDF and scale uncertainties from the theory are given as $\delta_{\mbox{\tiny stat}}$, $\delta_{\mbox{\tiny uncorr}}$, $\delta_{\mbox{\tiny corr}}$, $\delta_{\mbox{\tiny DPDF}}$ and $\delta_{\mbox{\tiny scal}}$, respectively.}
\label{table1Dratios}
}
\end{center}
\end{table}

\begin{table}[ht]
\begin{center}
{\footnotesize
\begin{tabular}{|ccc|c|c|c|c|c|c|}
\hline
&&&&&&&& \\[-0.3cm]
\multicolumn{3}{|c|}{$x^{\rm jets}_{\gamma}$}& $E_{T}^{\rm jet1} [\mbox{GeV}]$ & $\frac{\mbox{d}^{2}\sigma}{\mbox{d}x^{\rm jets}_{\gamma}\mbox{d}E_{T}^{\rm jet1}}\,\,[\mbox{pb}~\mbox{GeV}^{-1}]$ & $\delta_{\mbox{\tiny stat}}$ & $\delta_{\mbox{\tiny uncorr}}$ & $\delta_{\mbox{\tiny corrr}}$ & $1+\delta_{\mbox{\tiny hadr}}$ \\[0.4cm]
\hline
${\scriptstyle 0.1}$&\!\!\!\!${\scriptstyle \div}$\!\!\!\!&${\scriptstyle0.75}$&${\scriptstyle 5 \,\,\div  \,\, 6.5}$&${\scriptstyle 124}$&${\scriptstyle \pm 4}$&${\scriptstyle +12/-16}$ &  ${\scriptstyle +20/-19}$ & ${\scriptstyle 0.84}$\\
${\scriptstyle 0.75}$&\!\!\!\!${\scriptstyle \div}$\!\!\!\!&${\scriptstyle1}$&${\scriptstyle 5 \,\,\div  \,\, 6.5}$&${\scriptstyle 133}$&${\scriptstyle \pm 7}$&${\scriptstyle +13/-15}$ &  ${\scriptstyle +13/-14}$ & ${\scriptstyle 0.72}$\\
${\scriptstyle 0.1}$&\!\!\!\!${\scriptstyle \div}$\!\!\!\!&${\scriptstyle0.75}$&${\scriptstyle 6.5 \,\,\div \,\, 8.5}$&${\scriptstyle 40.2}$&${\scriptstyle \pm 1.9}$&${\scriptstyle +4.6/-4.8}$ &  ${\scriptstyle +6.4/-6.5}$ & ${\scriptstyle 0.99}$\\
${\scriptstyle 0.75}$&\!\!\!\!${\scriptstyle \div}$\!\!\!\!&${\scriptstyle1}$&${\scriptstyle 6.5 \,\,\div \,\, 8.5}$&${\scriptstyle 85}$&${\scriptstyle \pm 4}$&${\scriptstyle +9/-10}$ &  ${\scriptstyle +10/-10}$ & ${\scriptstyle 0.84}$\\
${\scriptstyle 0.1}$&\!\!\!\!${\scriptstyle \div}$\!\!\!\!&${\scriptstyle0.75}$&${\scriptstyle 8.5 \,\,\div \,\, 14}$&${\scriptstyle 2.60}$&${\scriptstyle \pm 0.25}$&${\scriptstyle +0.30/-0.39}$ &  ${\scriptstyle +0.64/-0.45}$ & ${\scriptstyle 1.0}$\\
${\scriptstyle 0.75}$&\!\!\!\!${\scriptstyle \div}$\!\!\!\!&${\scriptstyle1}$&${\scriptstyle 8.5 \,\,\div \,\, 14}$&${\scriptstyle 12.2}$&${\scriptstyle \pm 0.7}$&${\scriptstyle +1.3/-1.5}$ &  ${\scriptstyle +1.8/-1.8}$ & ${\scriptstyle 0.88}$\\

\hline
\hline
&&&&&&&& \\[-0.3cm]
\multicolumn{3}{|c|}{$z^{\rm jets}_{\pom}$}& $x^{\rm jets}_{\gamma}$ & $\frac{\mbox{d}^{2}\sigma}{\mbox{d}z^{\rm jets}_{\pom}\mbox{d}x^{\rm jets}_{\gamma}}~[\mbox{pb}]$ & $\delta_{\mbox{\tiny stat}}$ & $\delta_{\mbox{\tiny uncorr}}$ & $\delta_{\mbox{\tiny corrr}}$ & $1+\delta_{\mbox{\tiny hadr}}$ \\[0.4cm]
\hline
${\scriptstyle 0.14}$&\!\!\!\!${\scriptstyle \div}$\!\!\!\!&${\scriptstyle0.36}$&${\scriptstyle 0.1 \,\,\div \,\, 0.75}$&${\scriptstyle 403}$&${\scriptstyle \pm 23}$&${\scriptstyle +48/-59}$ &  ${\scriptstyle +76/-70}$ & ${\scriptstyle 0.91}$\\
${\scriptstyle 0.36}$&\!\!\!\!${\scriptstyle \div}$\!\!\!\!&${\scriptstyle0.58}$&${\scriptstyle 0.1 \,\,\div \,\, 0.75}$&${\scriptstyle 477}$&${\scriptstyle \pm 20}$&${\scriptstyle +57/-71}$ &  ${\scriptstyle +79/-72}$ & ${\scriptstyle 0.87}$\\
${\scriptstyle 0.58}$&\!\!\!\!${\scriptstyle \div}$\!\!\!\!&${\scriptstyle0.8}$&${\scriptstyle 0.1 \,\,\div \,\, 0.75}$&${\scriptstyle 386}$&${\scriptstyle \pm 16}$&${\scriptstyle +46/-53}$ &  ${\scriptstyle +59/-57}$ & ${\scriptstyle 0.87}$\\
${\scriptstyle 0.14}$&\!\!\!\!${\scriptstyle \div}$\!\!\!\!&${\scriptstyle0.36}$&${\scriptstyle 0.75 \,\,\div \,\, 1}$&${\scriptstyle 846}$&${\scriptstyle \pm 49}$&${\scriptstyle +10/-12}$ &  ${\scriptstyle +10/-11}$ & ${\scriptstyle 0.68}$\\
${\scriptstyle 0.36}$&\!\!\!\!${\scriptstyle \div}$\!\!\!\!&${\scriptstyle0.58}$&${\scriptstyle 0.75 \,\,\div \,\, 1}$&${\scriptstyle 605}$&${\scriptstyle \pm 32}$&${\scriptstyle +79/-81}$ &  ${\scriptstyle +61/-65}$ & ${\scriptstyle 0.85}$\\
${\scriptstyle 0.58}$&\!\!\!\!${\scriptstyle \div}$\!\!\!\!&${\scriptstyle0.8}$&${\scriptstyle 0.75 \,\,\div \,\, 1}$&${\scriptstyle 489}$&${\scriptstyle \pm 26}$&${\scriptstyle +55/-58}$ &  ${\scriptstyle +49/-51}$ & ${\scriptstyle 1.1}$\\

\hline
\end{tabular}
\caption[ab]{Bin averaged hadron level double-differential cross sections for diffractive dijet photoproduction. For each data point, the absolute statistical, uncorrelated and correlated systematic uncertainties and the hadronisation correction factors applied to the NLO calculations are given as $\delta_{\mbox{\tiny stat}}$, $\delta_{\mbox{\tiny uncorr}}$ and $\delta_{\mbox{\tiny corr}}$ and $1+\delta_{\mbox{\tiny hadr}}$, respectively.}
\label{table2D}
}
\end{center}
\end{table}

\begin{table}[ht]
\begin{center}
{\footnotesize
\begin{tabular}{|ccc|c|c|c|c|c|c|c|}
\hline
&&&&&&&&& \\[-10pt]\multicolumn{3}{|c|}{$E_{T}^{\rm jet1}\,\,[\mbox{GeV}]$}& $x^{\rm jets}_{\gamma}$ & $\mbox{data}/\mbox{theory}$ & $\delta_{\mbox{\tiny stat}}$ & $\delta_{\mbox{\tiny uncorr}}$ & $\delta_{\mbox{\tiny cororr}}$ & $\delta_{\mbox{\tiny DPDF}}$ & $\delta_{\mbox{\tiny scal}}$ \\[0.1cm]
\hline
${\scriptstyle 5}$&\!\!\!\!${\scriptstyle \div}$\!\!\!\!&${\scriptstyle 6.5}$&${\scriptstyle 0.1 \,\,\div \,\, 0.75}$&${\scriptstyle 0.52}$&${\scriptstyle \pm 0.02}$ &  ${\scriptstyle+0.05/-0.07}$ &  ${\scriptstyle+0.08/-0.08}$ &  ${\scriptstyle+0.09/-0.06}$ &  ${\scriptstyle+0.14/-0.13}$ \\
${\scriptstyle 6.5}$&\!\!\!\!${\scriptstyle \div}$\!\!\!\!&${\scriptstyle 8.5}$&${\scriptstyle 0.1 \,\,\div \,\, 0.75}$&${\scriptstyle 0.69}$&${\scriptstyle \pm 0.03}$ &  ${\scriptstyle+0.08/-0.08}$ &  ${\scriptstyle+0.11/-0.11}$ &  ${\scriptstyle+0.12/-0.07}$ &  ${\scriptstyle+0.24/-0.21}$ \\
${\scriptstyle 8.5}$&\!\!\!\!${\scriptstyle \div}$\!\!\!\!&${\scriptstyle 14}$&${\scriptstyle 0.1 \,\,\div \,\, 0.75}$&${\scriptstyle 0.78}$&${\scriptstyle \pm 0.08}$ &  ${\scriptstyle+0.09/-0.12}$ &  ${\scriptstyle+0.19/-0.13}$ &  ${\scriptstyle+0.14/-0.08}$ &  ${\scriptstyle+0.30/-0.25}$ \\
${\scriptstyle 5}$&\!\!\!\!${\scriptstyle \div}$\!\!\!\!&${\scriptstyle 6.5}$&${\scriptstyle 0.75 \,\,\div \,\, 1}$&${\scriptstyle 0.562}$&${\scriptstyle \pm 0.029}$ &  ${\scriptstyle+0.055/-0.065}$ &  ${\scriptstyle+0.054/-0.057}$ &  ${\scriptstyle+0.085/-0.050}$ &  ${\scriptstyle+0.073/-0.072}$ \\
${\scriptstyle 6.5}$&\!\!\!\!${\scriptstyle \div}$\!\!\!\!&${\scriptstyle 8.5}$&${\scriptstyle 0.75 \,\,\div \,\, 1}$&${\scriptstyle 0.63}$&${\scriptstyle \pm 0.03}$ &  ${\scriptstyle+0.07/-0.08}$ &  ${\scriptstyle+0.07/-0.07}$ &  ${\scriptstyle+0.11/-0.06}$ &  ${\scriptstyle+0.13/-0.12}$ \\
${\scriptstyle 8.5}$&\!\!\!\!${\scriptstyle \div}$\!\!\!\!&${\scriptstyle 14}$&${\scriptstyle 0.75 \,\,\div \,\, 1}$&${\scriptstyle 0.64}$&${\scriptstyle \pm 0.04}$ &  ${\scriptstyle+0.07/-0.08}$ &  ${\scriptstyle+0.10/-0.10}$ &  ${\scriptstyle+0.11/-0.07}$ &  ${\scriptstyle+0.15/-0.14}$ \\

\hline
\hline
&&&&&&&&& \\[-10pt]\multicolumn{3}{|c|}{$z^{\rm jets}_{\pom}$}& $x^{\rm jets}_{\gamma}$ & $\mbox{data}/\mbox{theory}$ & $\delta_{\mbox{\tiny stat}}$ & $\delta_{\mbox{\tiny uncorr}}$ & $\delta_{\mbox{\tiny cororr}}$ & $\delta_{\mbox{\tiny DPDF}}$ & $\delta_{\mbox{\tiny scal}}$ \\[0.1cm]
\hline
${\scriptstyle 0.1}$&\!\!\!\!${\scriptstyle \div}$\!\!\!\!&${\scriptstyle 0.75}$&${\scriptstyle 0.14 \,\,\div \,\, 0.36}$&${\scriptstyle 0.59}$&${\scriptstyle \pm 0.04}$ &  ${\scriptstyle+0.07/-0.09}$ &  ${\scriptstyle+0.11/-0.10}$ &  ${\scriptstyle+0.01/-0.05}$ &  ${\scriptstyle+0.14/-0.14}$ \\
${\scriptstyle 0.75}$&\!\!\!\!${\scriptstyle \div}$\!\!\!\!&${\scriptstyle 1}$&${\scriptstyle 0.14 \,\,\div \,\, 0.36}$&${\scriptstyle 0.666}$&${\scriptstyle \pm 0.038}$ &  ${\scriptstyle+0.079/-0.096}$ &  ${\scriptstyle+0.079/-0.084}$ &  ${\scriptstyle+0.110/-0.056}$ &  ${\scriptstyle+0.083/-0.093}$ \\
${\scriptstyle 0.1}$&\!\!\!\!${\scriptstyle \div}$\!\!\!\!&${\scriptstyle 0.75}$&${\scriptstyle 0.36 \,\,\div \,\, 0.58}$&${\scriptstyle 0.56}$&${\scriptstyle \pm 0.02}$ &  ${\scriptstyle+0.07/-0.08}$ &  ${\scriptstyle+0.09/-0.08}$ &  ${\scriptstyle+0.10/-0.05}$ &  ${\scriptstyle+0.16/-0.15}$ \\
${\scriptstyle 0.75}$&\!\!\!\!${\scriptstyle \div}$\!\!\!\!&${\scriptstyle 1}$&${\scriptstyle 0.36 \,\,\div \,\, 0.58}$&${\scriptstyle 0.546}$&${\scriptstyle \pm 0.029}$ &  ${\scriptstyle+0.071/-0.073}$ &  ${\scriptstyle+0.055/-0.059}$ &  ${\scriptstyle+0.084/-0.052}$ &  ${\scriptstyle+0.094/-0.087}$ \\
\hline
\end{tabular}
\caption{Ratios of double-differential cross sections for diffractive dijet photoproduction to NLO QCD calcualtions obtained in the FR framework with the H1 2006 Fit B DPDFs, corrected for hadronisation. For each ratio point, the absolute statistical, uncorrelated and correlated systematic uncertainties from the data and the absolute DPDF and scale uncertainties from the theory are given as $\delta_{\mbox{\tiny stat}}$,  $\delta_{\mbox{\tiny uncorr}}$, $\delta_{\mbox{\tiny corr}}$, $\delta_{\mbox{\tiny DPDF}}$  and $\delta_{\mbox{\tiny scal}}$, respectively.}
\label{table2Dratios}
}
\end{center}
\end{table}

\pagebreak
\begin{table}[ht]
\begin{center}
{\footnotesize
\begin{tabular}{|ccc|c|c|c|c|}
\hline
&&&&&&\\[-0.3cm]
\multicolumn{3}{|c|}{$x^{\rm jets}_{\gamma}$}& $\sigma^{\mbox{\tiny diff}}/\sigma^{\mbox{\tiny incl}}$ & $\delta_{\mbox{\tiny stat}}$ & $\delta_{\mbox{\tiny uncorr}}$ & $\delta_{\mbox{\tiny corr}}$ \\[0.1cm]
\hline
${\scriptstyle 0.1}$ &\!\!\!\!${\scriptstyle\div}$\!\!\!\!&${\scriptstyle 0.325}$&${\scriptstyle 0.00636}$ &  ${\scriptstyle \pm 0.00040}$ &  ${\scriptstyle +0.00063/-0.00069}$ &  ${\scriptstyle +0.00061/-0.00059}$ \\
${\scriptstyle 0.325}$ &\!\!\!\!${\scriptstyle\div}$\!\!\!\!&${\scriptstyle 0.55}$&${\scriptstyle 0.0157}$ &  ${\scriptstyle \pm 0.0007}$ &  ${\scriptstyle +0.0015/-0.0020}$ &  ${\scriptstyle +0.0013/-0.0013}$ \\
${\scriptstyle 0.55}$ &\!\!\!\!${\scriptstyle\div}$\!\!\!\!&${\scriptstyle 0.775}$&${\scriptstyle 0.0280}$ &  ${\scriptstyle \pm 0.0010}$ &  ${\scriptstyle +0.0030/-0.0026}$ &  ${\scriptstyle +0.0021/-0.0021}$ \\
${\scriptstyle 0.775}$ &\!\!\!\!${\scriptstyle\div}$\!\!\!\!&${\scriptstyle 1}$&${\scriptstyle 0.0353}$ &  ${\scriptstyle \pm 0.0012}$ &  ${\scriptstyle +0.0043/-0.0038}$ &  ${\scriptstyle +0.0027/-0.0028}$ \\
\hline
\hline
&&&&&&\\[-0.3cm]
\multicolumn{3}{|c|}{$E_{T}^{\rm jet1}\,\,[\!~\mbox{GeV}]$}& $\sigma^{\mbox{\tiny diff}}/\sigma^{\mbox{\tiny incl}}$ & $\delta_{\mbox{\tiny stat}}$ & $\delta_{\mbox{\tiny uncorr}}$ & $\delta_{\mbox{\tiny corr}}$ \\[0.1cm]
\hline
${\scriptstyle 5}$ &\!\!\!\!${\scriptstyle\div}$\!\!\!\!&${\scriptstyle 7}$&${\scriptstyle 0.0217}$ &  ${\scriptstyle \pm 0.0006}$ &  ${\scriptstyle +0.0020/-0.0020}$ &  ${\scriptstyle +0.0016/-0.0016}$ \\
${\scriptstyle 7}$ &\!\!\!\!${\scriptstyle\div}$\!\!\!\!&${\scriptstyle 10}$&${\scriptstyle 0.0126}$ &  ${\scriptstyle \pm 0.0005}$ &  ${\scriptstyle +0.0014/-0.0014}$ &  ${\scriptstyle +0.0010/-0.0010}$ \\
${\scriptstyle 10}$ &\!\!\!\!${\scriptstyle\div}$\!\!\!\!&${\scriptstyle 14}$&${\scriptstyle 0.00464}$ &  ${\scriptstyle \pm 0.00039}$ &  ${\scriptstyle +0.00070/-0.00054}$ &  ${\scriptstyle +0.00036/-0.00038}$ \\
\hline
\hline
&&&&&&\\[-0.3cm]
\multicolumn{3}{|c|}{$\left<\eta^{\rm jets}\right>$}& $\sigma^{\mbox{\tiny diff}}/\sigma^{\mbox{\tiny incl}}$ & $\delta_{\mbox{\tiny stat}}$ & $\delta_{\mbox{\tiny uncorr}}$ & $\delta_{\mbox{\tiny corr}}$ \\[0.1cm]
\hline
${\scriptstyle -0.7}$ &\!\!\!\!${\scriptstyle\div}$\!\!\!\!&${\scriptstyle -0.4}$&${\scriptstyle 0.0625}$ &  ${\scriptstyle \pm 0.0035}$ &  ${\scriptstyle +0.0067/-0.0063}$ &  ${\scriptstyle +0.0049/-0.0049}$ \\
${\scriptstyle -0.4}$ &\!\!\!\!${\scriptstyle\div}$\!\!\!\!&${\scriptstyle -0.1}$&${\scriptstyle 0.0437}$ &  ${\scriptstyle \pm 0.0016}$ &  ${\scriptstyle +0.0042/-0.0039}$ &  ${\scriptstyle +0.0033/-0.0033}$ \\
${\scriptstyle -0.1}$ &\!\!\!\!${\scriptstyle\div}$\!\!\!\!&${\scriptstyle 0.2}$&${\scriptstyle 0.0259}$ &  ${\scriptstyle \pm 0.0011}$ &  ${\scriptstyle +0.0028/-0.0030}$ &  ${\scriptstyle +0.0020/-0.0020}$ \\
${\scriptstyle 0.2}$ &\!\!\!\!${\scriptstyle\div}$\!\!\!\!&${\scriptstyle 0.5}$&${\scriptstyle 0.01525}$ &  ${\scriptstyle \pm 0.00070}$ &  ${\scriptstyle +0.0017/-0.0015}$ &  ${\scriptstyle 
+0.0011/-0.0011}$ \\
${\scriptstyle 0.5}$ &\!\!\!\!${\scriptstyle\div}$\!\!\!\!&${\scriptstyle 0.8}$&${\scriptstyle 0.00723}$ &  ${\scriptstyle \pm 0.00047}$ &  ${\scriptstyle +0.00078/-0.00084}$ &  ${\scriptstyle +0.00059/-0.00056}$ \\
\hline
\hline
&&&&&&\\[-0.3cm]
\multicolumn{3}{|c|}{$\left|\Delta \eta^{\rm jets}\right|$}& $\sigma^{\mbox{\tiny diff}}/\sigma^{\mbox{\tiny incl}}$ & $\delta_{\mbox{\tiny stat}}$ & $\delta_{\mbox{\tiny uncorr}}$ & $\delta_{\mbox{\tiny corr}}$ \\[0.1cm]
\hline
${\scriptstyle 0}$ &\!\!\!\!${\scriptstyle\div}$\!\!\!\!&${\scriptstyle 0.4}$&${\scriptstyle 0.0194}$ &  ${\scriptstyle \pm 0.0007}$ &  ${\scriptstyle +0.0019/-0.0015}$ &  ${\scriptstyle +0.0015/-0.0015}$ \\
${\scriptstyle 0.4}$ &\!\!\!\!${\scriptstyle\div}$\!\!\!\!&${\scriptstyle 0.8}$&${\scriptstyle 0.0197}$ &  ${\scriptstyle \pm 0.0008}$ &  ${\scriptstyle +0.0018/-0.0017}$ &  ${\scriptstyle +0.0015/-0.0015}$ \\
${\scriptstyle 0.8}$ &\!\!\!\!${\scriptstyle\div}$\!\!\!\!&${\scriptstyle 1.2}$&${\scriptstyle 0.0180}$ &  ${\scriptstyle \pm 0.0008}$ &  ${\scriptstyle +0.0014/-0.0018}$ &  ${\scriptstyle +0.0014/-0.0014}$ \\
${\scriptstyle 1.2}$ &\!\!\!\!${\scriptstyle\div}$\!\!\!\!&${\scriptstyle 1.6}$&${\scriptstyle 0.0148}$ &  ${\scriptstyle \pm 0.0008}$ &  ${\scriptstyle +0.0017/-0.0019}$ &  ${\scriptstyle +0.0011/-0.0011}$ \\
${\scriptstyle 1.6}$ &\!\!\!\!${\scriptstyle\div}$\!\!\!\!&${\scriptstyle 2}$&${\scriptstyle 0.0109}$ &  ${\scriptstyle \pm 0.0008}$ &  ${\scriptstyle +0.0012/-0.0011}$ &  ${\scriptstyle +0.0009/-0.0009}$ \\
\hline
\hline
&&&&&&\\[-0.3cm]
\multicolumn{3}{|c|}{$M_{X}\,\,[\!~\mbox{GeV}]$}& $\sigma^{\mbox{\tiny diff}}/\sigma^{\mbox{\tiny incl}}$ & $\delta_{\mbox{\tiny stat}}$ & $\delta_{\mbox{\tiny uncorr}}$ & $\delta_{\mbox{\tiny corr}}$ \\[0.1cm]
\hline
${\scriptstyle 9}$ &\!\!\!\!${\scriptstyle\div}$\!\!\!\!&${\scriptstyle 14}$&${\scriptstyle 0.0222}$ &  ${\scriptstyle \pm 0.0006}$ &  ${\scriptstyle +0.0019/-0.0020}$ &  ${\scriptstyle +0.0017/-0.0017}$ \\
${\scriptstyle 14}$ &\!\!\!\!${\scriptstyle\div}$\!\!\!\!&${\scriptstyle 19}$&${\scriptstyle 0.0159}$ &  ${\scriptstyle \pm 0.0006}$ &  ${\scriptstyle +0.0018/-0.0017}$ &  ${\scriptstyle +0.0012/-0.0012}$ \\
${\scriptstyle 19}$ &\!\!\!\!${\scriptstyle\div}$\!\!\!\!&${\scriptstyle 29}$&${\scriptstyle 0.0079}$ &  ${\scriptstyle \pm 0.0005}$ &  ${\scriptstyle +0.0010/-0.0007}$ &  ${\scriptstyle +0.0008/-0.0006}$ \\
\hline
\hline
&&&&&&\\[-0.3cm]
\multicolumn{3}{|c|}{$W\,\,[\!~\mbox{GeV}]$}& $\sigma^{\mbox{\tiny diff}}/\sigma^{\mbox{\tiny incl}}$ & $\delta_{\mbox{\tiny stat}}$ & $\delta_{\mbox{\tiny uncorr}}$ & $\delta_{\mbox{\tiny corr}}$ \\[0.1cm]
\hline
${\scriptstyle 175}$ &\!\!\!\!${\scriptstyle\div}$\!\!\!\!&${\scriptstyle 191}$&${\scriptstyle 0.0180}$ &  ${\scriptstyle \pm 0.0009}$ &  ${\scriptstyle +0.0019/-0.0023}$ &  ${\scriptstyle +0.0013/-0.0013}$ \\
${\scriptstyle 191}$ &\!\!\!\!${\scriptstyle\div}$\!\!\!\!&${\scriptstyle 207}$&${\scriptstyle 0.0177}$ &  ${\scriptstyle \pm 0.0007}$ &  ${\scriptstyle +0.0030/-0.0013}$ &  ${\scriptstyle +0.0014/-0.0015}$ \\
${\scriptstyle 207}$ &\!\!\!\!${\scriptstyle\div}$\!\!\!\!&${\scriptstyle 223}$&${\scriptstyle 0.0167}$ &  ${\scriptstyle \pm 0.0006}$ &  ${\scriptstyle +0.0014/-0.0021}$ &  ${\scriptstyle +0.0015/-0.0015}$ \\
${\scriptstyle 223}$ &\!\!\!\!${\scriptstyle\div}$\!\!\!\!&${\scriptstyle 239}$&${\scriptstyle 0.0159}$ &  ${\scriptstyle \pm 0.0008}$ &  ${\scriptstyle +0.0015/-0.0014}$ &  ${\scriptstyle +0.0013/-0.0012}$ \\
${\scriptstyle 239}$ &\!\!\!\!${\scriptstyle\div}$\!\!\!\!&${\scriptstyle 255}$&${\scriptstyle 0.0144}$ &  ${\scriptstyle \pm 0.0008}$ &  ${\scriptstyle +0.0014/-0.0014}$ &  ${\scriptstyle +0.0012/-0.0012}$ \\
\hline
\end{tabular}
\caption{{Ratios of the diffractive to the inclusive single-differential
hadron level cross sections.  The corresponding statistical, uncorrelated and correlated systematic uncertainties propagated to the ratio are given by $\delta_{\mbox{\tiny stat}}$, $\delta_{\mbox{\tiny uncorr}}$ and $\delta_{\mbox{\tiny corr}}$, respectively.}}
\label{tabledifftoincl}
}
\end{center}
\end{table}

\endgroup

\end{document}